\documentclass[manuscript, letterpaper]{aastex6}

\pdfoutput=1

\newcommand{\githash}{fd40d6a8}\newcommand{\gitdate}{2017-07-19}
\newcommand{\exampleindata}{200}
\newcommand{\exampleinwalkers}{32}
\newcommand{\exampleinburn}{500}
\newcommand{\exampleinsteps}{2000}
\newcommand{\exampleineff}{1737}
\newcommand{\exampleiindata}{100}
\newcommand{\exampleiinwalkers}{32}
\newcommand{\exampleiinburn}{500}
\newcommand{\exampleiinsteps}{2000}
\newcommand{\exampleiineff}{1134}
\newcommand{\exampleiiindata}{6950}
\newcommand{\exampleiiinwalkers}{32}
\newcommand{\exampleiiinburn}{500}
\newcommand{\exampleiiinsteps}{5000}
\newcommand{\exampleiiineff}{2900}

\newcommand{\exampleivnwalkers}{32}
\newcommand{\exampleivnburn}{5000}
\newcommand{\exampleivnsteps}{15000}
\newcommand{\exampleivneff}{1443}
\newcommand{\examplevndata}{1000}
\newcommand{\examplevnwalkers}{32}
\newcommand{\examplevnburn}{10000}
\newcommand{\examplevnsteps}{30000}
\newcommand{\examplevneff}{1490}

\newcommand{\rotationperiod}{\ensuremath{{3.80 \pm 0.15}}}

\usepackage{microtype}
\usepackage{url}
\usepackage{amsmath}
\usepackage{amssymb}
\usepackage{natbib}
\usepackage{multirow}
\makeatletter
\def\env@matrix{\hskip -\arraycolsep\let\@ifnextchar\new@ifnextchar\array{*{\c@MaxMatrixCols}c}}
\makeatother

\hypersetup{colorlinks = false}
\makeatletter 
\long\def\frontmatter@title@above{
\vspace*{-\headsep}\vspace*{\headheight}
\noindent\footnotesize
{\noindent\footnotesize\textsc{\@journalinfo}}\par
{\noindent\scriptsize Preprint typeset using \LaTeX\ style AASTeX6 with modifications
}\par\vspace*{-\baselineskip}\vspace*{0.625in}
}%
\makeatother

\makeatletter
\let\origsection\section
\renewcommand\section{\@ifstar{\starsection}{\nostarsection}}
\newcommand\nostarsection[1]{\sectionprelude\origsection{#1}}
\newcommand\starsection[1]{\sectionprelude\origsection*{#1}}
\newcommand\sectionprelude{\vspace{1em}}
\let\origsubsection\subsection
\renewcommand\subsection{\@ifstar{\starsubsection}{\nostarsubsection}}
\newcommand\nostarsubsection[1]{\subsectionprelude\origsubsection{#1}}
\newcommand\starsubsection[1]{\subsectionprelude\origsubsection*{#1}}
\newcommand\subsectionprelude{\vspace{1em}}
\makeatother

\sloppypar

\newcommand{\project}[1]{\textsf{#1}}
\newcommand{\kepler}{\project{Kepler}}
\newcommand{\lsst}{\project{LSST}}
\newcommand{\tess}{\project{TESS}}
\newcommand{\celerite}{\project{celerite}}
\newcommand{\celeriteterm}{\emph{celerite}}
\newcommand{\emcee}{\project{emcee}}

\newcommand{\foreign}[1]{\emph{#1}}

\newcommand{\etc}{\foreign{etc.}}

\newcommand{\figureref}[1]{\ref{fig:#1}}
\newcommand{\Figure}[1]{Figure~\figureref{#1}}
\newcommand{\figurelabel}[1]{\label{fig:#1}}

\renewcommand{\eqref}[1]{\ref{eq:#1}}
\newcommand{\Eq}[1]{Equation~(\eqref{#1})}
\newcommand{\eq}[1]{\Eq{#1}}
\newcommand{\eqalt}[1]{Equation~\eqref{#1}}
\newcommand{\eqlabel}[1]{\label{eq:#1}}

\newcommand{\sectionname}{Section}
\newcommand{\sectref}[1]{\ref{sect:#1}}
\newcommand{\Sect}[1]{\sectionname~\sectref{#1}}
\newcommand{\sect}[1]{\Sect{#1}}

\newcommand{\App}[1]{Appendix~\sectref{#1}}
\newcommand{\app}[1]{\App{#1}}
\newcommand{\sectlabel}[1]{\label{sect:#1}}

\newcommand{\T}{\ensuremath{\mathrm{T}}}
\newcommand{\dd}{\ensuremath{\,\mathrm{d}}}

\newcommand{\bvec}[1]{{\ensuremath{\boldsymbol{#1}}}}

\newcommand{\response}[1]{{#1}}

\begin{document}

\title{%
\vspace{-\baselineskip}
Fast and scalable Gaussian process modeling with applications
to astronomical time series
\vspace{-2\baselineskip}
}

\newcounter{affilcounter}

\setcounter{affilcounter}{1}

\edef \sagan {\arabic{affilcounter}}\stepcounter{affilcounter}
\altaffiltext{\sagan}{Sagan Fellow}

\edef \uw {\arabic{affilcounter}}\stepcounter{affilcounter}
\altaffiltext{\uw}{Astronomy Department, University of Washington,
                   Seattle, WA}

\edef \guggenheim {\arabic{affilcounter}}\stepcounter{affilcounter}
\altaffiltext{\guggenheim}{Guggenheim Fellow}

\edef \iis {\arabic{affilcounter}}\stepcounter{affilcounter}
\altaffiltext{\iis}{Department of Computational and Data Sciences,
                    Indian Institute of Science, Bangalore, India}

\edef \simons {\arabic{affilcounter}}\stepcounter{affilcounter}
\altaffiltext{\simons}{Simons Fellow}

\edef \columbia {\arabic{affilcounter}}\stepcounter{affilcounter}
\altaffiltext{\columbia}{Department of Astronomy, Columbia University,
                         New York, NY}

\author{%
    Daniel~Foreman-Mackey\altaffilmark{\sagan,\uw},
    Eric~Agol\altaffilmark{\uw,\guggenheim},
    Sivaram~Ambikasaran\altaffilmark{\iis}, and
    Ruth~Angus\altaffilmark{\simons,\columbia}
}

\begin{abstract}

The growing field of large-scale time domain astronomy requires methods for
probabilistic data analysis that are computationally tractable, even with
large datasets.
Gaussian Processes are a popular class of models used for this purpose but,
since the computational cost scales, in general, as the cube of the number of
data points, their application has been limited to small datasets.
In this paper, we present a novel method for Gaussian Process modeling in
one-dimension where the computational requirements scale linearly with the
size of the dataset.
We demonstrate the method by applying it to simulated and real astronomical
time series datasets.
These demonstrations are examples of probabilistic inference of stellar
rotation periods, asteroseismic oscillation spectra, and transiting planet
parameters.
The method exploits structure in the problem when the covariance function is
expressed as a mixture of complex exponentials, without requiring evenly
spaced observations or uniform noise.
This form of covariance arises naturally when the process is a mixture of
stochastically-driven damped harmonic oscillators~--~providing a physical
motivation for and interpretation of this choice~--~but we also demonstrate
that it can be a useful effective model in some other cases.
We present a mathematical description of the method and compare it to existing
scalable Gaussian Process methods.
The method is fast and interpretable, with a range of potential applications
within astronomical data analysis and beyond.
We provide well-tested and documented open-source implementations of this
method in \project{C++}, \project{Python}, and \project{Julia}.

\end{abstract}


\section{Introduction}

In the astrophysical literature, Gaussian Processes
\citep[GPs;][]{Rasmussen:2006} have been used to model stochastic variability
in light curves of stars \citep{Brewer:2009}, active galactic nuclei
\citep{Kelly:2014}, and the logarithmic flux of X-ray binaries
\citep{Uttley:2005}.
They have also been used as models for the cosmic microwave background
\citep{Bond:1987, Bond:1999, Wandelt:2003}, correlated instrumental noise
\citep{Gibson:2012}, and spectroscopic calibration \citep{Czekala:2017,
Evans:2015}.
While these models are widely applicable, their use has been limited, in
practice, by the computational cost and scaling.
The cost of computing a general GP likelihood scales as the cube of the number
of data points $\mathcal{O}(N^3)$ and in the current era of large time domain
surveys~--~with as many as $\sim10^{4-9}$ targets with $\sim10^{3-5}$
observations each~--~this scaling is prohibitive.

\response{Existing astronomical time series datasets have already reached the
limit where na\"ive application GP models is no longer tractable.
NASA's \kepler\ Mission \citep{Borucki:2010}, for example, measured light
curves with more than 60,000 observations each for about 190,000 stars.
Current and forthcoming surveys such as \project{K2} \citep{Howell:2014},
\tess\ \citep{Ricker:2014}, \lsst\ \citep{Ivezic:2008}, \project{WFIRST}
\citep{Spergel:2015}, and \project{PLATO} \citep{Rauer:2014} will continue to
produce similar or larger data volumes.}

In this paper, we present a method for directly and exactly computing a class
of GP models that scales linearly with the number of data points
$\mathcal{O}(N)$ for one dimensional data sets.
\response{Unlike earlier linear methods using Kalman filters \citep[for
example,][]{Kelly:2014}, this novel algorithm exploits the semiseparable
structure of a specific class of covariance matrices to directly factorize and
solve the system.
}
This method can only be used with one-dimensional datasets and the covariance
function must be represented by a mixture of exponentials; we will return to a
discussion of what this means in detail in the following sections.
However, the measurements don't need to be evenly spaced and the uncertainties
can be heteroscedastic.

This method achieves linear scaling by exploiting structure in the covariance
matrix when it is generated by a mixture of exponentials.
\response{The semiseparable nature of these matrices was first recognized by
\citet{Ambikasaran:2015}, building on intuition from a twenty year old paper
\citep{Rybicki:1995}.}
As we will discuss in the following pages, this choice of kernel function
arises naturally in physical systems and we demonstrate it can be used as
an effective model in other cases.
This method is especially appealing compared to other similar methods~--~we
discuss these comparisons in \sect{compare}~--~because it is exact, simple,
and fast.

Our main expertise lies in the field of exoplanet discovery and
characterization where GPs have become a model of choice.
We are confident that this method will benefit this field, but we also expect
that there will be applications in other branches of astrophysics and beyond.
In \sect{examples}, we present applications of the method to research problems
in stellar rotation (\sect{rotation}), asteroseismic analysis (\sect{astero}),
and exoplanet transit fitting (\sect{transit}).
Some readers might consider starting with these section for motivating
examples before delving into the detailed derivations of the earlier sections.

In the following pages, we motivate the general problem of GP modeling, derive
our novel direct solver, and demonstrate the model's application on real and
simulated data sets.
Alongside this paper, we have released well-tested and documented open source
implementations written in \project{C++}, \project{Python}, and
\project{Julia}.
These are available online at \project{GitHub}
\url{https://github.com/dfm/celerite} and \project{Zenodo}
\citep{Foreman-Mackey:2017}.

\section{Gaussian processes}\sectlabel{gps}

GPs are stochastic models consisting of a mean function
$\mu_\bvec{\theta}(\bvec{x})$ and a covariance, autocorrelation, or ``kernel''
function $k_\bvec{\alpha}(\bvec{x}_n,\,\bvec{x}_m)$ parameterized by
$\bvec{\theta}$ and $\bvec{\alpha}$ respectively.
Under this model, \response{the log-likelihood function $\mathcal{L}
(\bvec{\theta},\,\bvec{\alpha})$} with a dataset
\begin{eqnarray}
\bvec{y} = \left(\begin{array}{ccccc}
    y_1\quad && \cdots\quad && y_N
\end{array}\right)^\T
\end{eqnarray}
at coordinates
\begin{eqnarray}
X = \left(\begin{array}{ccccc}
    \bvec{x}_1\quad && \cdots\quad && \bvec{x}_N
\end{array}\right)^\T
\end{eqnarray}
is
\begin{eqnarray}\eqlabel{gp-likelihood}
\ln \mathcal{L} (\bvec{\theta},\,\bvec{\alpha}) =
\ln{p(\bvec{y}\,|\,{X,\,\bvec{\theta}},\,\bvec{\alpha})} =
    -\frac{1}{2} {\bvec{r}_\bvec{\theta}}^\T\,{K_\bvec{\alpha}}^{-1}\,
        \bvec{r}_\bvec{\theta}
    -\frac{1}{2}\ln\det K_\bvec{\alpha}
    - \frac{N}{2} \ln{(2\pi)}
\end{eqnarray}
where
\begin{eqnarray}
    \bvec{r}_\bvec{\theta} = \left(\begin{array}{ccccc}
    y_1 - \mu_\bvec{\theta}(\bvec{x}_1)\quad && \cdots\quad &&
    y_N - \mu_\bvec{\theta}(\bvec{x}_N)
\end{array}\right)^\T
\end{eqnarray}
is the vector of residuals and the elements of the covariance matrix $K$ are
given by $[K_\bvec{\alpha}]_{nm} = k_\bvec{\alpha}(\bvec{x}_n,\,\bvec{x}_m)$.
\eq{gp-likelihood} is the equation of an $N$-dimensional Gaussian and it can
be derived as the generalization of what we astronomers call ``$\chi^2$''
to include the effects of correlated noise.
\response{
A point estimate for the values of the parameters $\bvec{\theta}$ and
$\bvec{\alpha}$ for a given dataset $(\bvec{y},\,X)$ can be found by
maximizing \eq{gp-likelihood} with respect to $\bvec{\theta}$ and
$\bvec{\alpha}$ using a non-linear optimization routine \citep{Nocedal:2006}.}
\response{Furthermore, the uncertainties on $\bvec{\theta}$ and
$\bvec{\alpha}$ can be quantified by multiplying the likelihood by a prior
$p(\bvec{\theta},\,\bvec{\alpha})$ and using a Markov chain Monte Carlo (MCMC)
algorithm to sample the joint posterior probability density.}

GP models have been widely used across the physical sciences but their
application is generally limited to small datasets, \response{$N \lesssim
10^3$}, because the cost of computing the inverse and determinant of a general
matrix $K_\bvec{\alpha}$ is $\mathcal{O}(N^3)$.
In other words, this cost is proportional to the cube of the number of data
points $N$.
Furthermore, the storage requirements also scale as $\mathcal{O}(N^2)$.
This means that for large datasets, every evaluation of the likelihood quickly
becomes computationally intractable, and representing the matrix in memory can
become expensive.
In this case, the use of non-linear optimization or MCMC is no longer
practical.

\response{This is not a purely academic concern.
While the cost of directly evaluating the GP likelihood using a tuned linear
algebra library\footnote{For comparisons throughout this paper, we use the
Intel Math Kernel Library (\url{https://software.intel.com/en-us/intel-mkl})
running on a recent Apple MacBook Pro.} with a dataset of $N=1024$
measurements is less than a tenth of a second, the same calculation with
$N=8192$ takes over 8~seconds.
Furthermore, most optimization and sampling routines are iterative, requiring
many evaluations of this model to perform parameter estimation or inference.
Existing datasets from astronomical surveys like \kepler\ and \project{K2}
include tens of thousands of observations for hundreds of thousands of
targets, and upcoming surveys like \lsst\ are expected to produce thousands or
tens of thousands of measurements each for billions of astronomical sources
\citep{Ivezic:2008}.
Scalable methods will be required if we ever hope to use GP models with these
datasets.
}

In this paper, we present a method for improving the cubic scaling for some
use cases.
We call our method and its implementations \celerite.\footnote{The name
\celerite\ comes from the French word \foreign{c\'elerit\'e} meaning the speed
of light in a vacuum.
\response{Throughout the paper, when referring to the \celeriteterm\ model, we
typeset \celeriteterm\ in italics and, in reference to the implementation of
this model, we typeset \celerite\ in sans-serif.}}
The \celerite\ method requires using a specific model
for the covariance $k_\bvec{\alpha}(\bvec{x}_n,\,\bvec{x}_m)$ and, although it
has some limitations, we demonstrate in subsequent sections that it can
increase the computational efficiency of many astronomical data analysis
problems.
The main limitation of this method is that it can only be applied to
one-dimensional datasets, where by ``one-dimensional'' we mean that the
 \emph{input coordinates} $\bvec{x}_n$ are scalar, $\bvec{x}_n \equiv
t_n$.
We use $t$ as the input coordinate because one-dimensional GPs are often
applied to time series data but this isn't a real restriction and the
\celerite\ method can be applied to \emph{any} one-dimensional dataset.
Furthermore, the covariance function for the \celerite\ method is
``stationary''.
In other words, $k_\bvec{\alpha}(t_n,\,t_m)$ is only a function of $\tau_{nm}
\equiv |t_n - t_m|$.

\section{The celerite model}\sectlabel{celerite}

To scale GP models to larger datasets, \citet{Rybicki:1995} presented a method
of computing the first term in \eq{gp-likelihood} in $\mathcal{O}(N)$
operations when the covariance function is given by
\begin{eqnarray}\eqlabel{kernel-simple}
k_\bvec{\alpha}(\tau_{nm}) = \sigma_n^2\,\delta_{nm} + a\,\exp(-c\,\tau_{nm})
\end{eqnarray}
where $\{{\sigma_n}^2\}_{n=1}^N$ are the measurement uncertainties,
$\delta_{nm}$ is the Kronecker delta, and $\bvec{\alpha} = (a,\,c)$.
The intuition behind this method is that, for this choice of $k_\bvec{\alpha}$,
the inverse of $K_\bvec{\alpha}$ is tridiagonal and can be computed
with a small number of operations for each data point.
\response{This model has been generalized to arbitrary mixtures of
exponentials \citep[for example,][]{Kelly:2011}
\begin{eqnarray}
k_\bvec{\alpha}(\tau_{nm}) = \sigma_n^2\,\delta_{nm} +
    \sum_{j=1}^J a_j\,\exp(-c_j\,\tau_{nm}) \quad.
\end{eqnarray}
In this case, the inverse is dense but \eq{gp-likelihood} can still be
evaluated with a scaling of $\mathcal{O}(N\,J^2)$, where $J$ is the number of
components in the mixture \citep{Kelly:2014,Ambikasaran:2015}.
In \sect{factor}, we build on previous work \citep{Ambikasaran:2015} to derive
a faster method with the same $\mathcal{O}(N\,J^2)$ scaling that can be used
when $k_\bvec{\alpha}(\tau_{nm})$ is positive definite.}

This kernel function can be made even more general by
introducing complex parameters $a_j \to a_j\pm i\,b_j$ and
$c_j \to c_j\pm i\,d_j$.
In this case, the covariance function becomes
\begin{eqnarray}\eqlabel{celerite-kernel-complex}
k_\bvec{\alpha}(\tau_{nm}) = \sigma_n^2\,\delta_{nm} +
    \sum_{j=1}^J &&\left[
    \frac{1}{2}(a_j + i\,b_j)\,\exp\left(-(c_j+i\,d_j)\,\tau_{nm}\right)
        \right. \nonumber\\
    &&+\left.
    \frac{1}{2}(a_j - i\,b_j)\,\exp\left(-(c_j-i\,d_j)\,\tau_{nm}\right)
\right]
\end{eqnarray}
and, for this function, \eq{gp-likelihood} can still be evaluated with
$\mathcal{O}(N\,J^2)$ operations.
The details of this method and a few implementation considerations are
outlined in \sect{factor}, but we first discuss some properties of this
covariance function.

By rewriting the exponentials in \eq{celerite-kernel-complex} as sums of sine
and cosine functions, we can see the autocorrelation structure is defined by a
mixture of quasiperiodic oscillators
\begin{eqnarray}\eqlabel{celerite-kernel}
k_\bvec{\alpha}(\tau_{nm}) = \sigma_n^2\,\delta_{nm} +
    \sum_{j=1}^J &&\left[
    a_j\,\exp\left(-c_j\,\tau_{nm}\right)\,\cos\left(d_j\,\tau_{nm}\right)
        \right.\nonumber\\
    &&+ \left.
    b_j\,\exp\left(-c_j\,\tau_{nm}\right)\,\sin\left(d_j\,\tau_{nm}\right)
\right] \quad.
\end{eqnarray}
For clarity, we refer to the argument within the sum as a ``\celeriteterm\
term'' for the remainder of this paper.
The Fourier transform\footnote{Here and throughout we have defined the Fourier
transform of the function $f(t)$ as $F(\omega)={(2\,\pi)}^{-1/2}\,
\int_{-\infty}^\infty f(t)\,e^{i\,\omega\,t}\dd t$.} of this covariance
function is the power spectral density (PSD) of the process and it is given by
\begin{eqnarray}\eqlabel{celerite-psd}
S(\omega) = \sum_{j=1}^J \sqrt{\frac{2}{\pi}}
\frac{(a_j\,c_j+b_j\,d_j)\,({c_j}^2+{d_j}^2)+(a_j\,c_j-b_j\,d_j)\,\omega^2}
{\omega^4+2\,({c_j}^2-{d_j}^2)\,\omega^2+{({c_j}^2+{d_j}^2)}^2}\quad.
\end{eqnarray}
The physical interpretation of this model isn't immediately obvious and we
return to a more general discussion shortly but we start by considering some
special cases.

If we set the imaginary amplitude $b_j$ for some component $j$ to zero, that
term of \eq{celerite-kernel} becomes
\begin{eqnarray}
k_j(\tau_{nm}) =
    a_j\,\exp\left(-c_j\,\tau_{nm}\right)\,\cos\left(d_j\,\tau_{nm}\right)
\end{eqnarray}
and the PSD for this component is
\begin{eqnarray}\eqlabel{lorentz-psd}
S_j(\omega) = \frac{1}{\sqrt{2\,\pi}}\,\frac{a_j}{c_j}\,\left[
    \frac{1}{1+{\left(\frac{\omega-d_j}{c_j}\right)}^2} +
    \frac{1}{1+{\left(\frac{\omega+d_j}{c_j}\right)}^2}
\right] \quad.
\end{eqnarray}
This is the sum of two Lorentzian or Cauchy distributions with width $c_j$
centered on $\omega = \pm d_j$.
This model can be interpreted intuitively as a quasiperiodic oscillator with
amplitude $A_j = a_j$, quality factor $Q_j = d_j\,{(2\,c_j)}^{-1}$, and period
$P_j = 2\,\pi\,{d_j}^{-1}$.

Similarly, setting both $b_j$ and $d_j$ to zero, we get
\begin{eqnarray}
k_j(\tau_{nm}) = a_j\,\exp\left(-c_j\,\tau_{nm}\right)
\end{eqnarray}
with the PSD
\begin{eqnarray}
S_j(\omega) = \sqrt{\frac{2}{\pi}}\,\frac{a_j}{c_j}\,
    \frac{1}{1+{\left(\frac{\omega}{c_j}\right)}^2} \quad.
\end{eqnarray}
This model is often called an Ornstein--Uhlenbeck process \response{\citep[in
reference to the classic paper,][]{Uhlenbeck:1930}} and this is the kernel
that was studied by Rybicki \& Press \citep{Rybicki:1992,Rybicki:1995}.

Finally, we note that the product of two terms of the form found inside the
sum in \eq{celerite-kernel} can also be re-written as a sum with updated
parameters
\begin{eqnarray}\eqlabel{product-rule}
k_j(\tau) \, k_k(\tau) =
    e^{-\tilde{c}\,\tau}\,[
        \tilde{a}_+\,\cos(\tilde{d}_+\,\tau) + \tilde{b}_+\,\sin(\tilde{d}_+\,\tau) +
        \tilde{a}_-\,\cos(\tilde{d}_-\,\tau) + \tilde{b}_-\,\sin(\tilde{d}_-\,\tau)
    ]
\end{eqnarray}
where
\begin{eqnarray}
    \tilde{a}_{\pm} &=& \frac{1}{2}\,(a_j\,a_k \pm b_j\,b_k) \\
    \tilde{b}_{\pm} &=& \frac{1}{2}\,(b_j\,a_k \mp a_j\,b_k) \\
    \tilde{c} &=& c_j + c_k \\
    \tilde{d}_{\pm} &=& d_j \mp d_k \quad.
\end{eqnarray}
Therefore, the method described in \sect{factor} can be used to
perform scalable inference on large datasets for any model, where the kernel
function is a sum or product of \celeriteterm\ terms.

\section{celerite as a model of stellar variations}\sectlabel{sho}

We now turn to a discussion of \celeriteterm\ as a model of astrophysical
variability.
A common concern in the context of GP modeling in astrophysics is the lack of
physical motivation for the choice of kernel functions.
Kernels are often chosen simply because they are popular, with little
consideration of the impact of this decision.
In this section, we discuss an exact physical interpretation of the
\celeriteterm\ kernel that is applicable to many astrophysical systems, but
especially the time-variability of stars.

Many astronomical objects are variable on timescales determined by their
physical structure.
For phenomena such as stellar (asteroseismic) oscillations, variability is
excited by noisy physical processes and grows most strongly at the
characteristic timescale but is also damped due to dissipation in the system.
These oscillations are strong at resonant frequencies determined by the
internal stellar structure, which are both excited and damped by convective
turbulence.

To relate this physical picture to \celeriteterm, we consider the dynamics of a
stochastically-driven damped simple harmonic oscillator (SHO).
The differential equation for this system is
\begin{equation}
    \left[\frac{\dd^2}{\dd t^2} + \frac{\omega_0}{Q}\,\frac{\dd}{\dd t}
    + \omega_0^2\right]\, y(t) = \epsilon(t)
\end{equation}
where $\omega_0$ is the frequency of the undamped oscillator, $Q$ is the
quality factor of the oscillator, and $\epsilon(t)$ is a stochastic driving
force.
If $\epsilon(t)$ is white noise, the PSD of this process is
\citep{Anderson:1990}
\begin{equation}\eqlabel{sho-psd}
S(\omega) = \sqrt{\frac{2}{\pi}}\,\frac{S_0\,\omega_0^4}
    {(\omega^2-\omega_0^2)^2 + \omega_0^2\omega^2/Q^2}
\end{equation}
where $S_0$ is proportional to the power at $\omega = \omega_0$, $S(\omega_0)
= \sqrt{2/\pi}\,S_0\,Q^2$.
The power spectrum in \eq{sho-psd} matches \eq{celerite-psd} if
\begin{eqnarray}\eqlabel{sho-complex}
a_j &=& S_0\,\omega_0\,Q \\
b_j &=& \frac{S_0\,\omega_0\,Q}{\sqrt{4\,Q^2-1}} \nonumber\\
c_j &=& \frac{\omega_0}{2\,Q} \nonumber\\
d_j &=& \frac{\omega_0}{2\,Q} \sqrt{4\,Q^2-1} \quad, \nonumber
\end{eqnarray}
for $Q \ge \frac{1}{2}$.
For $0 < Q \le \frac{1}{2}$, \eq{sho-psd} can be captured by a pair of
\celeriteterm\
terms with parameters
\begin{eqnarray}\eqlabel{sho-real}
a_{j\pm} &=& \frac{1}{2}\,S_0\,\omega_0\,Q\,\left[ 1 \pm
        \frac{1}{\sqrt{1-4\,Q^2}}\right] \\
b_{j\pm} &=& 0 \nonumber\\
    c_{j\pm} &=& \frac{\omega_0}{2\,Q}\,\left[1 \mp \sqrt{1-4\,Q^2}\right]
    \nonumber\\
d_{j\pm} &=& 0 \quad. \nonumber
\end{eqnarray}

For these definitions, the kernel is
\begin{equation}\eqlabel{sho-kernel}
k_\mathrm{SHO}(\tau;\,S_0,\,Q,\,\omega_0) =
S_0\,\omega_0\,Q\,e^{-\frac{\omega_0\,\tau}{2Q}}\,
\begin{cases}
    \cosh{(\eta\,\omega_0\,\tau)} +
        \frac{1}{2\,\eta\,Q}\,\sinh{(\eta\,\omega_0\,\tau)}, & 0 < Q < 1/2\\
    2\,(1+\omega_0\,\tau), & Q = 1/2\\
    \cos{(\eta\,\omega_0\,\tau)} +
        \frac{1}{2\,\eta\,Q} \sin{(\eta\,\omega_0\,\tau)},& 1/2 < Q\\
\end{cases}
\end{equation}
where $\eta = \vert 1-(4\,Q^2)^{-1}\vert^{1/2}$.
Because of the damping, the characteristic oscillation frequency in this
model, $d_j$, for any finite quality factor $Q > 1/2$, is not equal to the
frequency of the undamped oscillator, $\omega_0$.

The power spectrum in \eq{sho-psd} has several limits of physical interest:
\begin{itemize}

{\item For $Q = 1/\sqrt{2}$, \eq{sho-psd} simplifies to
\begin{eqnarray}\eqlabel{granulation-psd}
S(\omega) = \sqrt{\frac{2}{\pi}}\,\frac{S_0}{(\omega/\omega_0)^4+1} \quad.
\end{eqnarray}
This functional form is commonly used to model for the background granulation
noise in asteoreseismic and helioseismic analyses \citep{Harvey:1985, Michel:2009,
Kallinger:2014}.
The kernel for this value of $Q$ is
\begin{equation}
k(\tau) = S_0\,\omega_0\,e^{-\frac{1}{\sqrt{2}}\,\omega_0\,\tau}\,
    \cos{\left(\frac{\omega_0\,\tau}{\sqrt{2}}-\frac{\pi}{4}\right)} \quad.
\end{equation}}

{\item Substituting $Q = 1/2$, \eq{sho-psd} becomes
\begin{eqnarray}
S(\omega) =
    \sqrt{\frac{2}{\pi}}\,\frac{S_0}{\left[(\omega/\omega_0)^2+1\right]^2}
\end{eqnarray}
with the corresponding covariance function (using
\eqalt{celerite-kernel} and \eqalt{sho-real})
\begin{eqnarray}\eqlabel{approx-matern}
k(\tau) &=& \lim_{f \to 0}\,
    \frac{1}{2}\,S_0\,\omega_0\,
    \left[\left(1+1/f\right)\,e^{-\omega_0\,(1-f)\,\tau} +
          \left(1-1/f\right)\,e^{-\omega_0\,(1+f)\,\tau}
    \right] \\
&=& S_0\,\omega_0\,e^{-\omega_0\,\tau}\,[1+\omega_0\,\tau]
\end{eqnarray}
or, equivalently (using \eqalt{celerite-kernel} and \eqalt{sho-complex})
\begin{eqnarray}\eqlabel{approx-matern2}
k(\tau) &=& \lim_{f \to 0}\,
    S_0\,\omega_0\,e^{-\omega_0\,\tau}\,
    \left[\cos(f\,\tau) + \frac{\omega_0}{f}\,\sin(f\,\tau)\right] \\
&=& S_0\,\omega_0\,e^{-\omega_0\,\tau}\,[1+\omega_0\,\tau] \quad.
\end{eqnarray}
This covariance function is also known as the Mat\'ern-3/2 function
\citep{Rasmussen:2006}.
\response{This model cannot be directly evaluated using \celerite\ because,
as we can see from \eq{sho-complex}, the parameter $b_j$ is infinite for $Q =
1/2$.
The limit in \eq{approx-matern2}, however, suggests that an approximate
Mat\'ern-3/2 covariance can be computed with \celerite\ by using a small value
of $f$ in \eq{approx-matern2}.
The required value of $f$ will depend on the specific dataset and the precision
requirements for the inference, so we encourage users to test this
approximation in the context of their research before using it.
}}

{\item Finally, in the limit of large $Q$, the model approaches a high
    quality oscillation with frequency $\omega_0$ and covariance function
\begin{eqnarray}
k(\tau) \approx
    S_0\,\omega_0\,Q\,
    \exp\left(-\frac{\omega_0\,\tau}{2\,Q}\right)\,
    \cos\left(\omega_0\,\tau\right) \quad.
\end{eqnarray}}

\end{itemize}
\Figure{sho} shows a plot of the PSD for these limits and several other values
of $Q$.
From this figure, it is clear that for $Q \le 1/2$, the model has no
oscillatory behavior and that for large $Q$, the shape of the PSD near the
peak frequency approaches a Lorentzian.

These special cases demonstrate that the stochastically-driven simple harmonic
oscillator provides a physically motivated model that is flexible enough to
describe a wide range of stellar variations and we return to give some
specific examples in \sect{examples}.
Low $Q \approx 1$ can capture granulation noise and high $Q \gg 1$ is a good
model for asteroseismic oscillations.
In practice, we take a sum over oscillators with different values of $Q$,
$S_0$, and $\omega_0$ to give a sufficient accounting of the power spectrum
of stellar time series.
Since this kernel is exactly described by the exponential kernel, the
likelihood (\eqalt{gp-likelihood}) can be evaluated for a time series with $N$
measurements in $\mathcal{O}(N)$ operations using the \celeriteterm\ method
described in the next section.

\begin{figure}[!hptb]
\begin{center}
\includegraphics[width=0.95\textwidth]{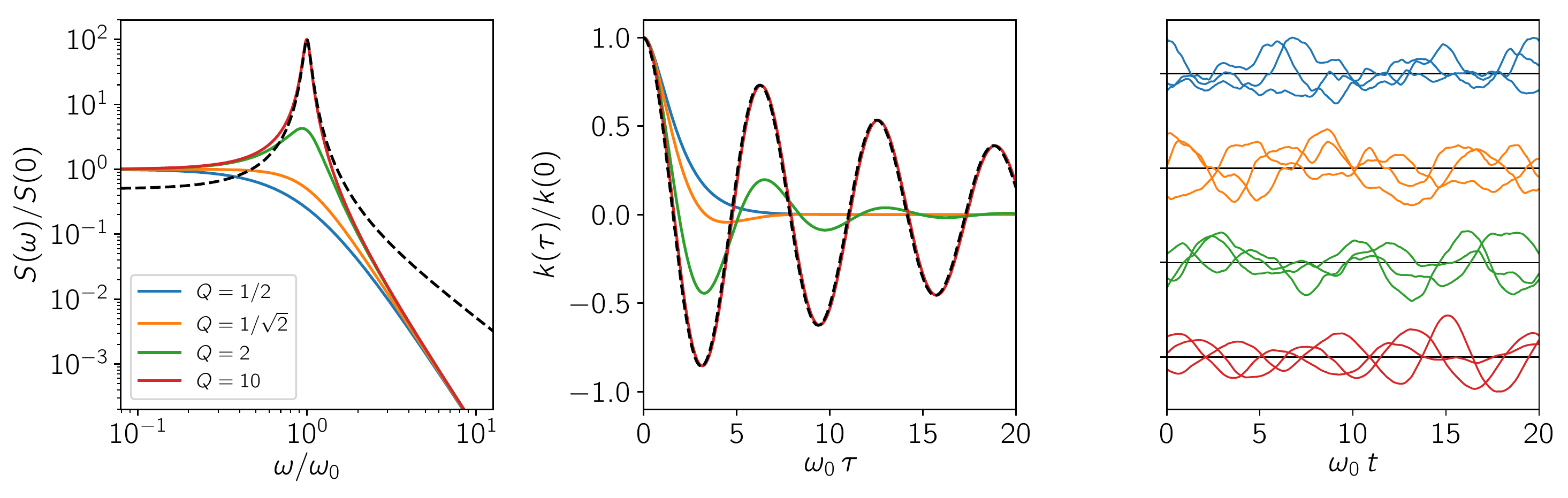}
\caption{(left) The power spectrum of a stochastically-driven simple harmonic
    oscillator (\eqalt{sho-psd}) plotted for several values of the quality
    factor $Q$.
    For comparison, the dashed line shows the Lorentzian function from
    \eq{lorentz-psd} with $c_j = \omega_0/(2\,Q) = 1/20$ and normalized so that
    $S(d_j)/S(0) = 100$.
    (middle) The corresponding autocorrelation functions with the same colors.
    \response{(right) Three realizations in arbitrary units from each model in
    the same colors.}
    \figurelabel{sho}}
\end{center}
\end{figure}

\response{
\section{Semiseparable matrices \& celerite}\sectlabel{factor}

The previous two sections describe the \celeriteterm\ model and its physical
interpretation.
In this section, we derive a new scalable direct solver for covariance
matrices generated by this model to compute the GP likelihood
(\eqalt{gp-likelihood}) for large datasets.
This method exploits the semiseparable structure of these matrices to compute
their Cholesky factorization in $\mathcal{O}(N\,J^2)$ operations.

The relationship between the \celeriteterm\ model and semiseparable linear
algebra was first recognized by Ambikasaran \citep{Ambikasaran:2015} building
on earlier work by Rybicki \& Press \citep{Rybicki:1995}.
Ambikasaran derived a scalable direct solver for any general semiseparable
matrix and applied this method to \celeriteterm\ models with $b_j = 0$ and
$d_j = 0$.
During the preparation of this paper, we generalized this earlier method to
include non-zero $b_j$ and $d_j$, but it turns out that we can derive a higher
performance algorithm by restricting our method to positive definite matrices.
In \app{psd}, we discuss methods for ensuring that a \celeriteterm\ model is
positive definite.

To start, we define a rank-$R$ semiseparable matrix as a matrix $K$ where the
elements can be written as
\begin{eqnarray}
K_{n,m} = \left\{\begin{array}{ll}
    \sum_{r=1}^R U_{n,r}\,V_{m,r} & \mathrm{if}\,m<n \\
    A_{n,n} & \mathrm{if}\,m=n \\
    \sum_{r=1}^R U_{m,r}\,V_{n,r} & \mathrm{otherwise}
\end{array}\right.
\end{eqnarray}
for some $N \times R$-dimensional matrices $U$ and $V$, and an $N \times
N$-dimensional diagonal matrix $A$.
This definition can be equivalently written as
\begin{eqnarray}\eqlabel{semi-sep}
K = A + \mathrm{tril}(U\,V^\T) + \mathrm{triu}(V\,U^\T)
\end{eqnarray}
where the $\mathrm{tril}$ function extracts the strict lower triangular
component of its argument and $\mathrm{triu}$ is the equivalent strict upper
triangular operation.
In this representation, rather than storing the full $N \times N$ matrix $K$,
it is sufficient to only store the $N \times R$ matrices $U$ and $V$, and the
$N$ diagonal entries of $A$, for a total storage volume of $(2\,R+1)\,N$
floating point numbers.
Given a semiseparable matrix of this form, the matrix-vector products and
matrix-matrix products can be evaluated in $\mathcal{O}(N)$ operations.
We include a summary of the key operations in \app{operations}, but the
interested reader is directed to \citet{Vandebril:2007} for more details about
semiseparable matrices in general.

\subsection{Cholesky factorization of positive definite semiseparable matrices}
\sectlabel{cholesky}

For our purposes, we want to compute the log-determinant of $K$ and solve
linear systems of the form $K\,\bvec{z}=\bvec{y}$ for $\bvec{z}$, where $K$ is
a rank-$R$ semiseparable positive definite matrix.
These can both be computed using the Cholesky factorization of $K$:
\begin{eqnarray}
K = L\,D\,L^\T
\end{eqnarray}
where $L$ is a lower triangular matrix with unit diagonal and $D$ is a
diagonal matrix.
We use the ansatz that $L$ has the form
\begin{eqnarray}
    L = I + \mathrm{tril} (U\,W^T)
\end{eqnarray}
where $I$ is the identity, $U$ is the matrix from above, and $W$ is an unknown
$N \times R$ matrix.
Combining this with \eq{semi-sep}, we find the equation
\begin{eqnarray}\eqlabel{cholesky}
A + \mathrm{tril}(U\,V^\T) + \mathrm{triu}(V\,U^\T)
&=& \left[I + \mathrm{tril} (U\,W^T)\right]\,D\,
    {\left[I + \mathrm{tril} (U\,W^T)\right]}^\T \\
&=& D + \mathrm{tril}(U\,W^\T)\,D
    + D\,\mathrm{triu}(W\,U^\T) \nonumber\\
&& +\, \mathrm{tril}(U\,W^\T)\,D\,\mathrm{triu}(W\,U^\T) \nonumber
\end{eqnarray}
that we can solve for $W$.
Setting each element on the left side of \eq{cholesky} equal to the
corresponding element on the right side, we derive the following recursion
relations
\begin{eqnarray}
S_{n,j,k} &=& S_{n-1,j,k} + D_{n-1,n-1}\,W_{n-1,j}\,W_{n-1,k} \nonumber\\
D_{n,n} &=& A_{n,n} -
    \sum_{j=1}^{R}\sum_{k=1}^{R} U_{n,j}\,S_{n,j,k}\,U_{n,k}
    \nonumber\\
W_{n,j} &=& \frac{1}{D_{n,n}}\left[ V_{n,j} -
    \sum_{k=1}^{R}U_{n,k}\,S_{n,j,k} \right]
    \eqlabel{algorithm}
\end{eqnarray}
where $S_n$ is a symmetric $R \times R$ matrix and every element
$S_{1,j,k}=0$.
The computational cost of one iteration of the update in \eq{algorithm} is
$\mathcal{O}(R^2)$ and this must be run $N$ times~--~once for each row~--~so
the total cost of this factorization scales as $\mathcal{O}(N\,R^2)$.

After computing the Cholesky factorization of this matrix $K$, we can also
apply its inverse and compute its log-determinant in $\mathcal{O}(N\,R)$ and
$\mathcal{O}(N)$ operations respectively.
The log-determinant of $K$ is given by
\begin{eqnarray}
    \ln \det K = \sum_{n=1}^N \ln D_{n,n}
\end{eqnarray}
where, since $K$ is positive definite, $D_{n,n}>0$ for all $n$.

The inverse of $K$ can be applied using the relationship
\begin{eqnarray}
\bvec{z} = K^{-1}\,\bvec{y} = {(L^\T)}^{-1}\,D^{-1}\,L^{-1}\,\bvec{y} \quad.
\end{eqnarray}
First, we solve for $\bvec{z}^\prime = L^{-1}\,\bvec{y}$ using forward
substitution
\begin{eqnarray}\eqlabel{fwd}
    f_{n,j} &=& f_{n-1,j} + W_{n-1,j}\,z_{n-1}^\prime \nonumber\\
    z_n^\prime &=& y_n - \sum_{j=1}^{R} U_{n,j}\,f_{n,j}
\end{eqnarray}
where $f_{0,j} = 0$ for all $j$.
Then, using this result and backward substitution, we compute $\bvec{z}$
\begin{eqnarray}\eqlabel{bwd}
g_{n,j} &=& g_{n+1,j} + U_{n+1,j}\,z_{n+1} \nonumber\\
z_n &=& \frac{z_n^\prime}{D_{n,n}} - \sum_{j=1}^{R} W_{n,j}\,g_{n,j}
\end{eqnarray}
where $g_{N+1,j} = 0$ for all $j$.
The total cost of the forward and backward substitution passes scales as
$\mathcal{O}(N\,R)$.

The algorithm derived here is a general method for factorizing, solving, and
computing the determinant of positive definite rank-$R$ semiseparable
matrices.
As we will show below, this method can be used to compute
\eq{gp-likelihood} for \celeriteterm\ models.
In practice, when $K$ is positive definite and this method is applicable, we
find that this method is about a factor of 20 faster than an optimized
implementation of the earlier, more general method from
\citet{Ambikasaran:2015}.

\subsection{Scalable computation of celerite models}\sectlabel{scalable}

The \celeriteterm\ model described in \sect{celerite} can be exactly
represented as a rank $R = 2\,J$ semiseparable matrix\footnote{We note that if
some \celeriteterm\ terms are real ($b_j=0$ and $d_j=0$) then these terms only
add one to the rank.
Therefore, if a \celeriteterm\ model has $J_\mathrm{real}$ real and
$J_\mathrm{complex}$ general terms, the semiseparable rank is only
$R=J_\mathrm{real} + 2\,J_\mathrm{complex}=2\,J-J_{real}$.}
where the components of the relevant matrices are
\begin{eqnarray}\eqlabel{system}
U_{n,2\,j-1} &=& a_j\,e^{-c_j\,t_n}\,\cos(d_j\,t_n) +
    b_j\,e^{-c_j\,t_n}\,\sin(d_j\,t_n) \nonumber\\
U_{n,2\,j} &=& a_j\,e^{-c_j\,t_n}\,\sin(d_j\,t_n) -
    b_j\,e^{-c_j\,t_n}\,\cos(d_j\,t_n) \nonumber\\
V_{m,2\,j-1} &=& e^{c_j\,t_m}\,\cos(d_j\,t_m) \nonumber\\
V_{m,2\,j} &=& e^{c_j\,t_m}\,\sin(d_j\,t_m) \nonumber\\
A_{n,n} &=& \sigma_n^2 + \sum_{j=1}^J a_j \quad.
\end{eqnarray}
A na\"ive implementation of the algorithm in \eq{algorithm} for this system
will result in numerical overflow and underflow issues in many realistic cases
because both $U$ and $V$ have factors of the form $e^{\pm c_j t}$, where $t$
can be any arbitrarily large real number.
This issue can be avoided by reparameterizing the semiseparable representation
from \eq{system}.
We define the following pre-conditioned variables
\begin{eqnarray}\eqlabel{reparam}
\tilde{U}_{n,2\,j-1} &=& a_j\,\cos(d_j\,t_n) + b_j\,\sin(d_j\,t_n) \nonumber\\
\tilde{U}_{n,2\,j} &=& a_j\,\sin(d_j\,t_n) - b_j\,\cos(d_j\,t_n) \nonumber\\
\tilde{V}_{m,2\,j-1} &=& \cos(d_j\,t_m) \nonumber\\
\tilde{V}_{m,2\,j} &=& \sin(d_j\,t_m) \nonumber\\
\tilde{W}_{n,2\,j-1} &=& e^{-c_j\,t_n}\,W_{n,2\,j-1} \nonumber\\
\tilde{W}_{n,2\,j} &=& e^{-c_j\,t_n}\,W_{n,2\,j}
\end{eqnarray}
and the new set of variables
\begin{eqnarray}
\phi_{n,2\,j-1} = \phi_{n,2\,j} = e^{-c_j\,(t_n - t_{n-1})}
\end{eqnarray}
with the constraint
\begin{eqnarray}
\phi_{1,2\,j-1} = \phi_{1,2\,j} = 0 \quad.
\end{eqnarray}
Using this parameterization, we find the following, numerically stable,
algorithm to compute the Cholesky factorization of a \celeriteterm\ model
\begin{eqnarray}\eqlabel{cholesky-factor}
    S_{n,j,k} &=& \phi_{n,j}\,\phi_{n,k}\,\left[S_{n-1,j,k} +
    D_{n-1,n-1}\,\tilde{W}_{n-1,j}\,\tilde{W}_{n-1,k}\right] \nonumber\\
D_{n,n} &=& A_{n,n} -
    \sum_{j=1}^{2\,J}\sum_{k=1}^{2\,J} \tilde{U}_{n,j}\,S_{n,j,k}\,\tilde{U}_{n,k}
    \nonumber\\
\tilde{W}_{n,j} &=& \frac{1}{D_{n,n}}\left[ \tilde{V}_{n,j} -
    \sum_{k=1}^{2\,J}\tilde{U}_{n,k}\,S_{n,j,k} \right] \quad.
\end{eqnarray}
As before, a single iteration of this algorithm requires $\mathcal{O}(J^2)$
operations and it is repeated $N$ times for a total scaling of
$\mathcal{O}(N\,J^2)$.
Furthermore, $\tilde{V}_{n,j}$ and $A_{n,n}$ can be updated in place and used
to represent $\tilde{W}_{n,j}$ and $D_{n,n}$, giving a total memory footprint
of $(6\,J + 1)\,N + J\,(J-1)/2$ floating point numbers.

The log-determinant of $K$ can still be computed as in the previous section,
but we must update the forward and backward substitution algorithms to account
for the reparameterization in \eq{reparam}.
The forward substitution with the reparameterized variables is
\begin{eqnarray}
    f_{n,j} &=& \phi_{n,j}\,\left[f_{n-1,j} +
    \tilde{W}_{n-1,j}\,z_{n-1}^\prime\right] \nonumber\\
    z_n^\prime &=& y_n - \sum_{j=1}^{2\,J} \tilde{U}_{n,j}\,f_{n,j}
\end{eqnarray}
where $f_{0,j} = 0$ for all $j$, and the backward substitution algorithm
becomes
\begin{eqnarray}
g_{n,j} &=& \phi_{n+1,j}\,\left[g_{n+1,j} +
    \tilde{U}_{n+1,j}\,z_{n+1}\right] \nonumber\\
z_n &=& \frac{z_n^\prime}{D_{n,n}} - \sum_{j=1}^{2\,J} \tilde{W}_{n,j}\,g_{n,j}
\end{eqnarray}
where $g_{N+1,j} = 0$ for all $j$.
The only differences, when compared to Equations~(\eqref{fwd})
and~(\eqref{bwd}) is the multiplication by $\phi_{n,j}$ in the recursion
relations for $f_{n,j}$ and $g_{n,j}$.
As before, the computational cost of this solve scales as $\mathcal{O}(N\,J)$.

\subsection{Performance}

We have derived a novel and scalable method for computing the Cholesky
factorization of the covariance matrices produced by \celeriteterm\ models.
In this section, we demonstrate the numerical and computational performance of
this algorithm.

First, we empirically confirm that the method solves systems of the form
$K\,\bvec{z} = \bvec{y}$ to high numerical accuracy.
The left panel of \Figure{error} shows the L-infinity norm for the residual
$\bvec{z} - K^{-1}\,\bvec{y} = \bvec{z}-K^{-1}K\bvec{z}$ for a range of dataset sizes $N$ and numbers of terms $J$.
For each pair of $N$ and $J$ we repeated the experiment 10 times with
different randomly generated datasets and \celeriteterm\ parameters, and
averaged across experiments to remove some sampling noise from this figure.
The numerical error introduced by our method increases weakly for larger
datasets $\sim N^{0.15}$ and roughly linearly with $J$.

For the experiments with small $N \le 2048$, we also compared the log
determinant calculation to the result calculated using the general log
determinant function provided by the \project{NumPy} project
\citep{Van-Der-Walt:2011}, itself built on the \project{LAPACK}
\citep{Anderson:1999} implementation in the Intel Math Kernel
Library\footnote{\url{https://software.intel.com/en-us/intel-mkl}}.
The right-hand panel of \Figure{error} shows a histogram of the fractional
error in the log determinant calculation for each of the 480~experiments with
$N \le 2048$.
These results agree to about $10^{-15}$ and the error only weakly increases
with $N$ and $J$.

We continue by measuring the real-world computational efficiency of this
method as a function of $N$ and $J$.
This experiment was run on a 2013 MacBook Pro with a dual 2.6 GHz Intel Core
i5 processor, but we find similar performance on other platforms.
The \celerite\ implementation used for this test is the \project{Python}
interface with a \project{C++} back end, but there is little overhead
introduced by \project{Python} so we find similar performance using
the implementation in \project{C++} and an independent implementation in
\project{Julia} \citep{Bezanzon:2012}.
\Figure{benchmark} shows the computational cost of evaluating
\eq{gp-likelihood} using \celerite\ as a function of $N$ and $J$.
The empirical scaling as a function of data size matches the theoretical
scaling of $\mathcal{O}(N)$ for all values of $N$ and the scaling with the
number of terms is the theoretical $\mathcal{O}(J^2)$ for $J \gtrsim 10$, with
some overhead for smaller models.
For comparison, the computational cost of the same computation using the Intel
Math Kernel Library, a highly optimized general dense linear algebra package,
is shown as a dashed line, which is independent of $J$.
\celerite\ is faster than the reference implementation for all but the
smallest datasets with many terms, $J \ga 64$ for $N\approx 512$,
and $J$ increasing with larger $N$.

\begin{figure}[tp]
\begin{center}
\includegraphics[width=\textwidth]{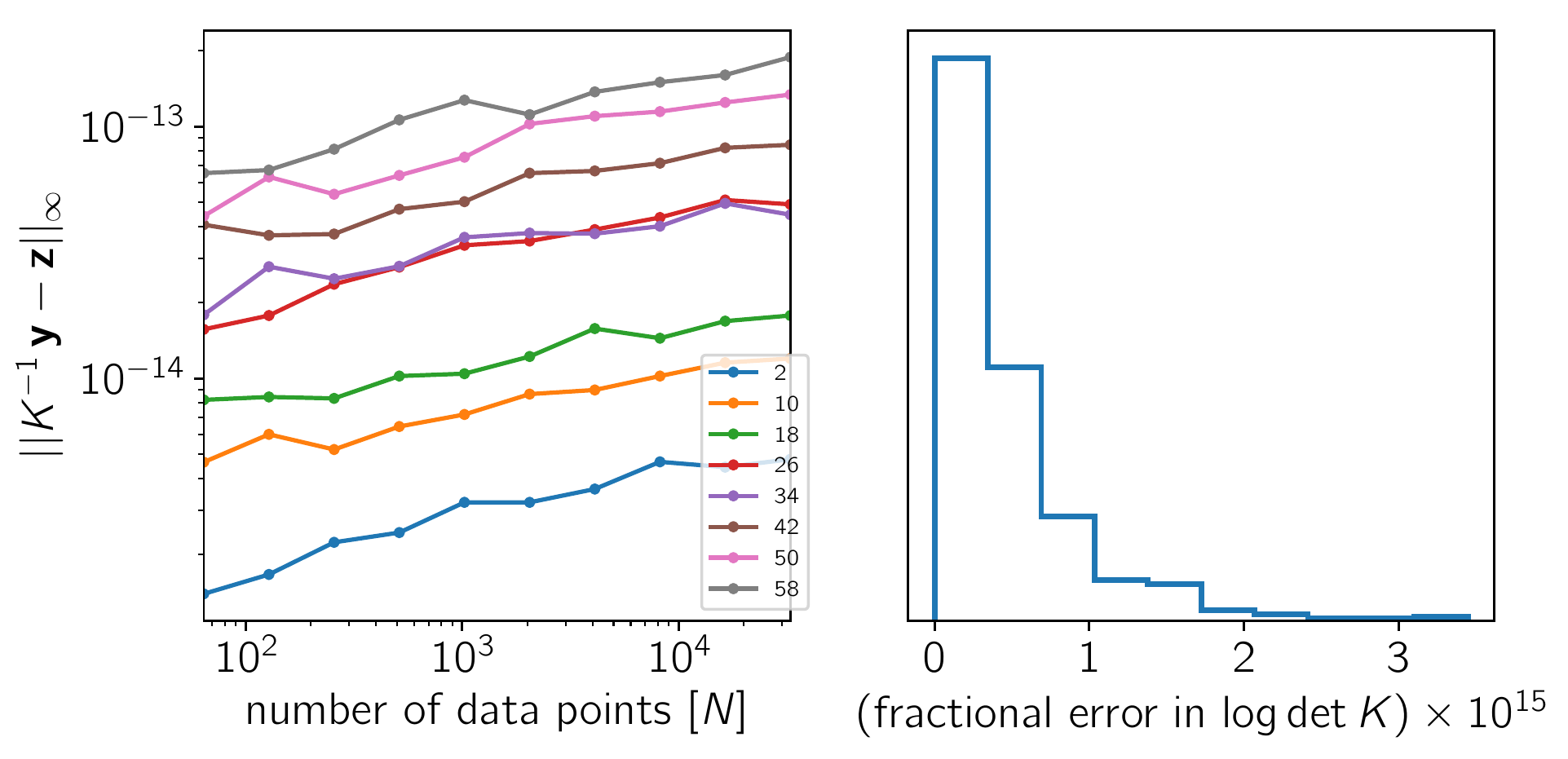}
\caption{\emph{(left)} The maximum numerical error introduced by the algorithm
    described in the text as a function of the number of \celeriteterm\ terms
    $J$ and data points $N$.
    Each line corresponds to a different $J$ and the error increases with $J$
    as shown in the legend.
    Each point in this figure is the average result across 10 systems with
    randomly sampled parameters.
\emph{(right)} The distribution of fractional error on the log determinant of
    $K$ computed using the algorithm described in the text compared to the
    algorithm implemented in \project{NumPy} \citep{Van-Der-Walt:2011} for the
    same systems shown in the left-hand figure.
    This histogram only includes comparisons for $N \le 2048$ because standard
    methods become computationally expensive for larger systems.
    \figurelabel{error}}
\end{center}
\end{figure}

\begin{figure}[tp]
\begin{center}
\includegraphics[width=\textwidth]{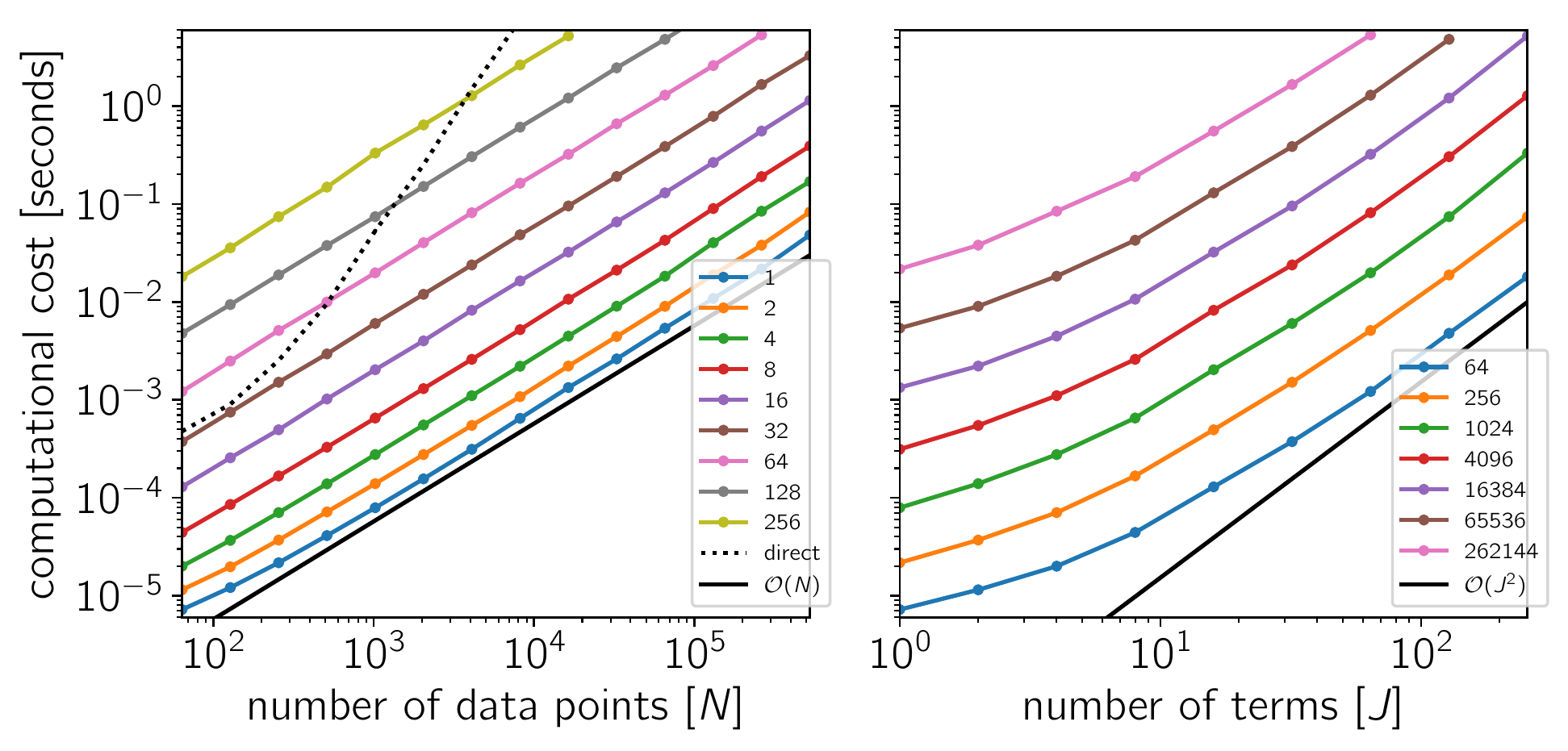}
\caption{A benchmark showing the computational scaling of \celerite\ using the
    Cholesky factorization routine for rank-$2\,J$ semiseparable matrices as
    described in the text.
    \emph{(left)} The cost of computing \eq{gp-likelihood} with a covariance
    matrix given by \eq{celerite-kernel} as a function of the number of data
    points $N$.
    The different lines show the cost for different numbers of terms $J$
    increasing from bottom to top.
    To guide the eye, the straight black line without points
    shows linear scaling in $N$.
    For comparison, the computational cost for the same model using a general
    Cholesky factorization routine implemented in the Intel MKL linear algebra
    package is shown as the dashed black line.
    \emph{(right)} The same information plotted as a function of $J$ for
    different values of $N$.
    Each line shows the scaling for a specific value of $N$ increasing from
    bottom to top.
    The black line shows quadratic scaling in $J$.
    \figurelabel{benchmark}}
\end{center}
\end{figure}

\newpage
\subsection{How to choose a celerite model}

In any application of \celerite~--~or any other GP model, for that
matter~--~we must choose a kernel.
A detailed discussion of model selection is beyond the scope of this paper and
the interested reader is directed to Chapter~5 of \citet{Rasmussen:2006} for
an introduction in the context of GPs, but we include a summary and some
suggestions specific to \celerite\ here.

For many of the astrophysics problems that we will tackle using \celerite, it
is possible to motivate the kernel choice physically.
We give three examples of this in the next section and, in these cases, we
have a physical description of the system that generated our data, we
represent that system as a \celeriteterm\ model, and use \celerite\ for
parameter estimation.
For other projects, we might be interested in comparing different theories
and, in those case, we would need to perform formal model selection.

Sometimes, it is not possible to choose a physically motivated kernel when we
use a GP as an effective model.
In these cases, we recommend starting with a mixture of stochastically-driven
damped SHOs (as discussed in \sect{sho}) and select the number of oscillators
using Bayesian model comparison, cross-validation, or another model selection
technique.
Despite the fact that it is less flexible than a general \celeriteterm\
kernel, we recommend the SHO model in most cases because it is once mean
square differentiable
\begin{eqnarray}
\left.\frac{\dd k_\mathrm{SHO}(\tau)}{\dd \tau}\right|_{\tau=0} = 0 \quad,
\end{eqnarray}
giving smoother functions than the general \celeriteterm\ model that is only
mean square continuous.

For a similar class of models~--~we return to these below in
\sect{compare}~--~\citet{Kelly:2014} recommend using an ``information
criterion'', such as the Akaike information criterion (AIC) or the Bayesian
information criterion (BIC), as a model selection metric.
These criteria are easy to implement and we use the BIC in \sect{example2}
but, since these are only formally valid under specific sets of assumptions, we
cannot recommend their use in general.
}

\section{Examples}\sectlabel{examples}

\response{
To demonstrate the application of \celerite, we use it to perform posterior
inference in two examples with realistic simulated data and three examples of
real-world research problems.
These examples have been chosen to justify the use of \celerite\ for a range
of research problems, but they are by no means exhaustive.
These examples are framed in the time domain with a clear bias in favor of
large homogeneous photometric surveys, but \celerite\ can also be used for
other one-dimensional problems like, for example, spectroscopy, where
wavelength~--~instead of time~--~is the independent coordinate \citep[see][for
a potential application]{Czekala:2017}.

In the first example, we demonstrate that \celerite\ can be used to infer
the power spectrum of a process when the data are generated from a
\celeriteterm\ model.
In the second example, we demonstrate that \celerite\ can be used as an
effective model even if the true process cannot be represented in the space of
allowed models.
This is an interesting example because, when analyzing real data, we rarely
have any fundamental reason to believe that the data were generated by a GP
model with a specific kernel.
Even in these cases, GPs can be useful effective models and \celerite\
provides computational advantages over other GP methods.

The next three examples show \celerite\ used to infer the properties of stars
and exoplanets observed by NASA's \kepler\ Mission.
Each of these examples touches on an active area of research so we limit our
examples to be qualitative in nature and do not claim that \celerite\ is the
optimal method, but we hope these examples encourage interested readers to
investigate the applicability of \celerite\ to their research.

In each example, the joint posterior probability density is given by
\begin{eqnarray}\eqlabel{joint-post}
p(\bvec{\theta},\,\bvec{\alpha}\,|\,\bvec{y},\,X) \propto
p(\bvec{y}\,|\,{X,\,\bvec{\theta}},\,\bvec{\alpha})\,
p(\bvec{\theta},\,\bvec{\alpha})
\end{eqnarray}
where $p(\bvec{y}\,|\,{X,\,\bvec{\theta}},\,\bvec{\alpha})$ is the GP
likelihood defined in \eq{gp-likelihood} and $p(\bvec{\theta},\,\bvec{\alpha})$
is the joint prior probability density for the parameters.
We sample each posterior density using MCMC and investigate the performance of
the model when used to interpret the physical processes that generated the
data.
The specific parameters and priors are discussed in detail in each section,
but we generally assume separable uniform priors with plausibly broad support.
}

\subsection{Example 1: Recovery of a celerite process}\sectlabel{example1}

In this first example, we simulate a dataset using a known \celeriteterm\
process and fit it with \celerite\ to show that valid inferences can be made
in this idealized case.
\response{We simulate a small ($N = \exampleindata$) dataset using a GP model
with a SHO kernel (\eqalt{sho-kernel}) with parameters $S_0 =
1\,\mathrm{ppm}^2$, $\omega_0 = e^2\,\mathrm{day}^{-1}$, and $Q = e^2$, where
the units are arbitrarily chosen to simplify the discussion.
We add further white noise with the amplitude $\sigma_n = 2.5\, \mathrm{ppm}$
to each data point.
The simulated data are plotted in the top left panel of
\Figure{simulated-correct}.

In this case, when simulating and fitting, we set the mean function
$\mu_\bvec{\theta}$ to zero~--~this means that the parameter vector
$\bvec{\theta}$ is empty~--~and the elements of the covariance matrix are
given by \eq{sho-kernel} with three parameters $\bvec{\alpha} = (S_0,\,
\omega_0,\,Q)$.
We choose a proper separable prior for $\bvec{\alpha}$ with log-uniform
densities for each parameter as listed in Table~\ref{tab:example-1-params}.

To start, we estimate the maximum \emph{a posteriori} (MAP) parameters using
the \project{L-BFGS-B} non-linear optimization routine \citep{Byrd:1995,
Zhu:1997} implemented by the \project{SciPy} project \citep{Jones:2001}.
The top left panel of \Figure{simulated-correct} shows the conditional mean
and standard deviation of the MAP model over-plotted on the simulated data and
the bottom panel shows the residuals away from this model.
Since we are using the correct model to fit the data, it is reassuring that
the residuals appear qualitatively uncorrelated in this figure.

We then sample the joint posterior probability (\eqalt{joint-post}) using
\emcee\ \citep{Goodman:2010, Foreman-Mackey:2013}.
We initialize \exampleinwalkers~walkers by sampling from a three-dimensional
isotropic Gaussian centered the MAP parameter vector and with a standard
deviation of $10^{-4}$ in each dimension.
We then run \exampleinburn~steps of burn-in and \exampleinsteps~steps of MCMC.
To assess convergence, we estimate the mean integrated autocorrelation time of
the chain across parameters \citep{Sokal:1989, Goodman:2010} and find that the
chain results in \exampleineff~effective samples.

Each sample in the chain corresponds to a model PSD and we compare this
posterior constraint on the PSD to the true spectral density in the
right-hand panel of \Figure{simulated-correct}.
In this figure, the true PSD is plotted as a dashed black line and the
numerical estimate of the posterior constraint on the PSD is shown as a blue
contour indicating 68\% of the MCMC samples.
It is clear from this figure that, as expected, the inference correctly
reproduces the true PSD.
}

\begin{floattable}
\begin{deluxetable}{cc}
\tablecaption{The parameters and priors for Example 1. \label{tab:example-1-params}}
\tablehead{\colhead{parameter} & \colhead{prior}}
\startdata
$\ln(S_0)$ & $\mathcal{U}(-10,\,10)$ \\
$\ln(Q)$ & $\mathcal{U}(-10,\,10)$ \\
$\ln(\omega_0)$ & $\mathcal{U}(-10,\,10)$ \\
\enddata
\end{deluxetable}
\end{floattable}

\begin{figure}[!hptb]
\begin{center}
\includegraphics[width=0.9\textwidth]{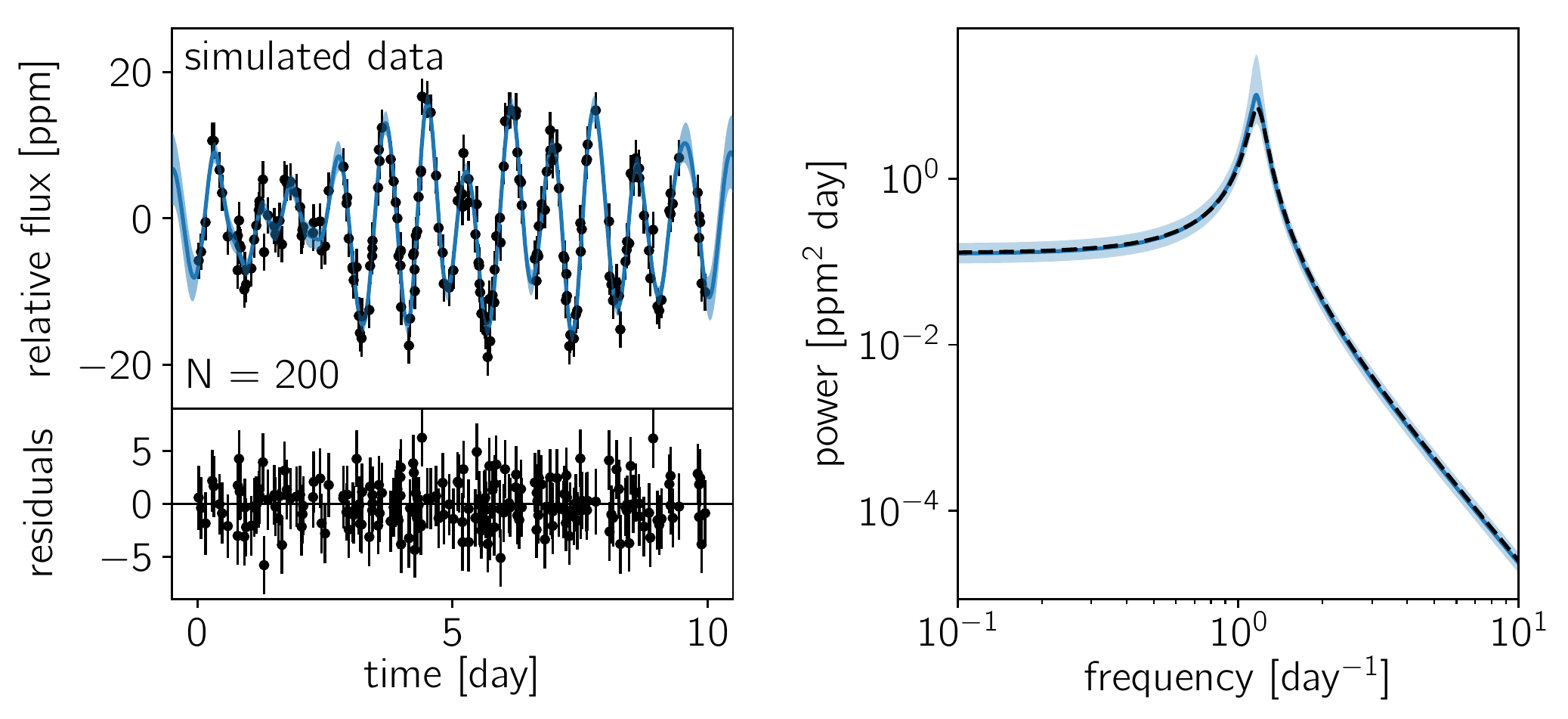}
\caption{(top left) A simulated dataset (black error bars), and MAP
    model (blue contours).
    (bottom left) The residuals between the mean predictive model and the data
    shown in the top left figure.
    (right) The inferred PSD~--~the blue contours encompass 68\% of the
    posterior mass~--~compared to the true PSD (dashed black line).
    \figurelabel{simulated-correct}}
\end{center}
\end{figure}

\subsection{Example 2: Inferences with the ``wrong'' model}\sectlabel{example2}

In this example, we simulate a dataset using a known GP model with a kernel
outside of the support of a \celeriteterm\ process.
This means the true autocorrelation of the process can never be correctly
represented by the model that we are using to fit, but we use this example
to demonstrate that, at least in this case, valid inferences can still be made
about the physical parameters of the model.

The data are simulated from a quasiperiodic GP with the kernel
\begin{eqnarray}\eqlabel{sim-wrong-true}
k_\mathrm{true} (\tau) = A\,
    \exp\left(-\frac{\tau^2}{2\,\lambda^2}\right)\,
    \cos\left(\frac{2\,\pi\,\tau}{P_\mathrm{true}}\right)
\end{eqnarray}
where $P_\mathrm{true}$ is the fundamental period of the process.
This autocorrelation structure corresponds to the power spectrum
\citep{Wilson:2013}
\begin{eqnarray}
S_\mathrm{true} (\omega) = \frac{\lambda\,A}{2}\,\left[
    \exp\left(-\frac{\lambda^2}{2}\,\left(\omega-
        \frac{2\,\pi}{P_\mathrm{true}}\right)^2\right) +
    \exp\left(-\frac{\lambda^2}{2}\,\left(\omega+
        \frac{2\,\pi}{P_\mathrm{true}}\right)^2\right)
\right]
\end{eqnarray}
which, for large values of $\omega$, falls off exponentially.
When compared to \eq{celerite-psd}~--~which, for large $\omega$, goes as
$\omega^{-4}$ at most~--~it is clear that a \celeriteterm\ model can never
exactly reproduce the structure of this process.
That being said, we demonstrate that robust inferences can be made about
$P_\mathrm{true}$ even with this effective model.

\response{We generate a realization of a GP model with the kernel given in
\eq{sim-wrong-true} with $N = \exampleiindata$, $A = 1\,\mathrm{ppm}^2$,
$\lambda = 5\,\mathrm{day}$, and $P_\mathrm{true} = 1\,\mathrm{day}$.
We also add white noise with amplitude $\sigma = 0.5\,\mathrm{ppm}$ to each
data point.
The top left panel of \Figure{simulated-wrong} shows this simulated dataset.

We then fit this simulated data using the product of two SHO terms
(\eqalt{sho-kernel}) where one of the terms has $S_0 = 1$ and $Q =
1/\sqrt{2}$ fixed.
The kernel for this model is
\begin{eqnarray}
k(\tau) = k_\mathrm{SHO}(\tau;\,S_0,\,Q,\,\omega_1) \,
    k_\mathrm{SHO}(\tau;\,S_0 = 1,\,Q = 1/\sqrt{2},\,\omega_2)
\end{eqnarray}
where $k_\mathrm{SHO}$ is defined in \eq{sho-kernel} and the period of the
process is $P=2\,\pi/\omega_1$.
We note that using \eq{product-rule}, the product of two \celeriteterm\ terms
can also be expressed using \celerite.

We choose this functional form by comparing the Bayesian information criterion
(BIC) for a set of different models.
In each model, we take the sum or product of $J$ SHO terms, find the maximum
likelihood model using \project{L-BFGS-B} and compute the BIC
\citep{Schwarz:1978}
\begin{eqnarray}
\mathrm{BIC} = -2\,\mathcal{L}^* + K\,\log N
\end{eqnarray}
where $\mathcal{L}^*$ is the value of the likelihood at maximum, $K$ is the
number of parameters, and $N$ is the number of data points.
\Figure{bic} shows the value of the BIC for a set of products and sums of SHO
terms and the model that we choose has the lowest BIC.

As in the previous example, we set the mean function to zero and can,
therefore, omit the parameters $\bvec{\theta}$.
Table~\ref{tab:example-2-params} lists the proper log-uniform priors that we
choose for each parameter in $\bvec{\alpha} =
(S_0,\,Q,\,\omega_1,\,\omega_2)$.
These priors, together with the GP likelihood (\eqalt{gp-likelihood}) fully
specify the posterior probability density.

As above, we estimate the MAP parameters using \project{L-BFGS-B} and sample
the posterior probability density using \emcee.
The top left panel of \Figure{simulated-wrong} shows the conditional mean and
standard deviation of the MAP model.
The bottom left panel shows the residuals between the data and this MAP model
and, even though this GP model is formally ``wrong'', there are no obvious
correlations in these residuals.

To perform posterior inference, we initialize \exampleiinwalkers~walkers by
sampling from the 4-dimensional Gaussian centered on the MAP parameters with
an isotropic standard deviation of $10^{-4}$.
We then run \exampleiinburn~steps of burn-in and \exampleiinsteps~steps of
MCMC.
To estimate the number of independent samples, we estimate the integrated
autocorrelation time of the chain for the parameter $\omega_1$~--~the
parameter of primary interest~--~and find \exampleiineff~effective samples.

For comparison, we run the same number of steps of MCMC to sample the
``correct'' joint posterior density.
For this reference inference, we use a GP likelihood with the kernel given by
\eq{sim-wrong-true} and choose log-uniform priors on each of the
three parameters $\ln A/\mathrm{ppm}^2 \sim \mathcal{U}(-10,\,10)$,
$\ln \lambda/\mathrm{day} \sim \mathcal{U}(-10,\,10)$, and
$\ln P_\mathrm{true}/\mathrm{day} \sim \mathcal{U}(-10,\,10)$.

The marginalized posterior inferences of the characteristic period of the
process are shown in the right panel of \Figure{simulated-wrong}.
The inference using the correct model is shown as a dashed blue histogram and
the inference made using the effective model is shown as a solid black
histogram.
These inferences are consistent with each other and with the true period used
for the simulation (shown as a vertical gray line).
This demonstrates that, in this case, \celerite\ can be used as a
computationally efficient effective model and hints that this may be true in
other problems as well.
}

\begin{floattable}
\begin{deluxetable}{cc}
\tablecaption{The parameters and priors for Example 2. \label{tab:example-2-params}}
\tablehead{\colhead{parameter} & \colhead{prior}}
\startdata
$\ln(S_0)$ & $\mathcal{U}(-15,\,5)$ \\
$\ln(Q)$ & $\mathcal{U}(-10,\,10)$ \\
$\ln(\omega_1)$ & $\mathcal{U}(-10,\,10)$ \\
$\ln(\omega_2)$ & $\mathcal{U}(-5,\,5)$ \\
\enddata
\end{deluxetable}
\end{floattable}

\begin{figure}[!hptb]
\begin{center}
\includegraphics[width=0.9\textwidth]{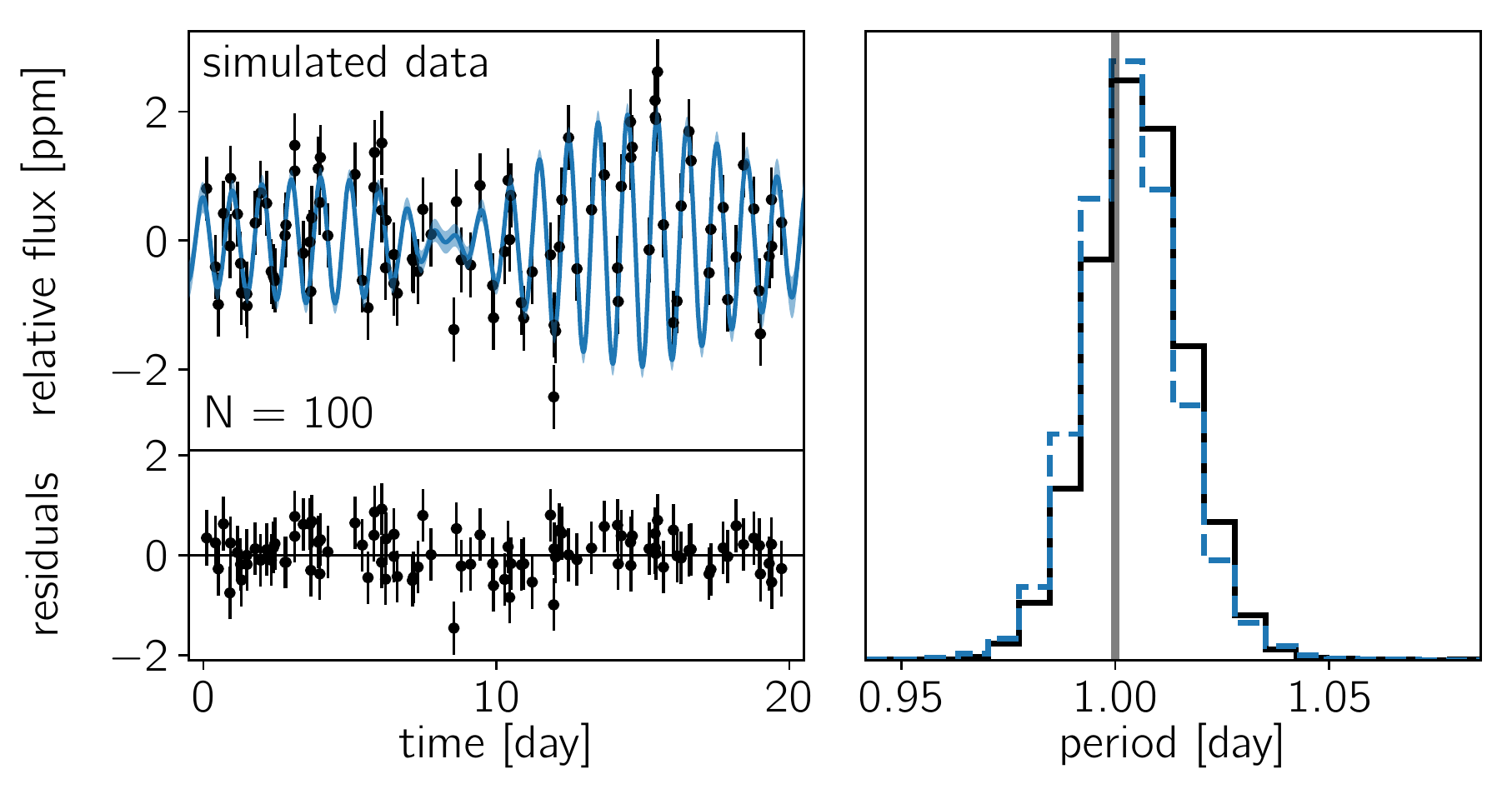}
\caption{(top left) A simulated dataset (black error bars), and MAP
    model (blue contours).
    (bottom left) The residuals between the mean predictive model and the data
    shown in the top left figure.
    (right) The inferred period of the process. The true period is indicated
    by the vertical orange line, the posterior inference using the correct
    model is shown as the blue dashed histogram, and the inference made using
    the ``wrong'' effective model is shown as the black histogram.
    \figurelabel{simulated-wrong}}
\end{center}
\end{figure}

\begin{figure}[!hptb]
\begin{center}
\includegraphics[width=0.5\textwidth]{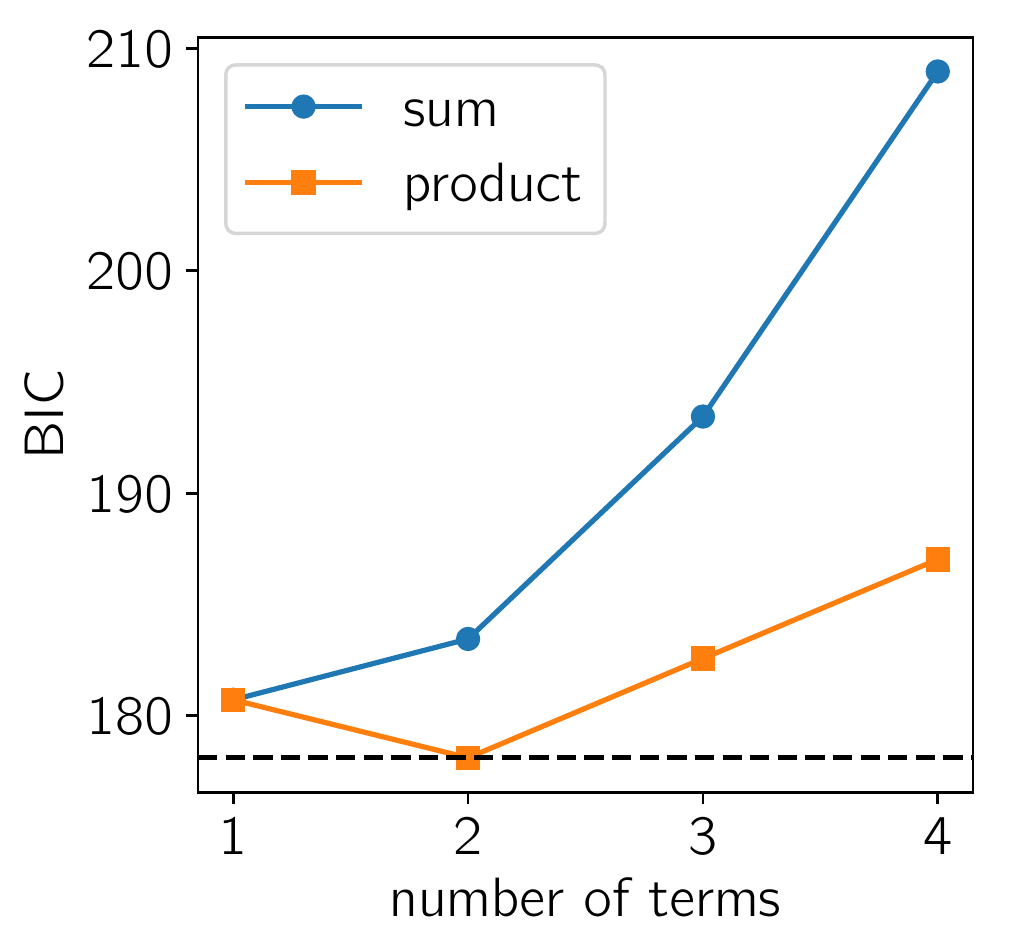}
\caption{\response{
    The Bayesian information criterion (BIC) evaluated for the data from
    \Figure{simulated-wrong} and different kernel choices.
    The $x$-axis shows the number of terms included in each model.
    The blue circles show models where the terms have been summed and the
    orange squares indicate that the model is a product of terms.
    In the product models, the parameters $S_0$ and $Q$ are held fixed at $1$
    and $1/\sqrt{2}$ respectively for all but the first term.
    The dashed black line shows the minium value of the BIC and this
    corresponds to the model chosen in the text.}
    \figurelabel{bic}}
\end{center}
\end{figure}

\subsection{Example 3: Stellar rotation} \sectlabel{rotation}

A source of variability that can be measured from time series measurements of
stars is rotation.
The inhomogeneous surface of the star (spots, plage, \etc) imprints itself as
quasiperiodic variations in photometric or spectroscopic observations
\citep{Dumusque:2014}.
It has been demonstrated that for light curves with nearly uniform sampling,
the empirical autocorrelation function provides a reliable estimate of the
rotation period of a star \citep{Mcquillan:2013, Mcquillan:2014, Aigrain:2015}
and that a GP model with a
quasiperiodic covariance function can be used to make probabilistic
measurements even with sparsely sampled data \citep{Angus:2017}.
The covariance function used for this type of analysis has the form
\begin{eqnarray}\eqlabel{sine2}
k(\tau) = A\,\exp\left(-\frac{\tau^2}{2\,\ell^2} -
    \Gamma\,\sin^2\left(\frac{\pi\,\tau}{P_\mathrm{rot}} \right) \right)
\end{eqnarray}
where $P_\mathrm{rot}$ is the rotation period of the star.
GP modeling with the same kernel function has been proposed as a method of
measuring the mean periodicity in quasiperiodic photometric time series in
general \citep{Wang:2012}.
The key difference between \eq{sine2} and other quasiperiodic kernels is that
it is positive for all values of $\tau$.
We construct a simple \celeriteterm\ covariance function with similar
properties as follows
\begin{eqnarray}\eqlabel{rot-kernel}
k(\tau) = \frac{B}{2+C}\,e^{-\tau/L}\,\left[
    \cos\left(\frac{2\,\pi\,\tau}{P_\mathrm{rot}}\right) + (1 + C)
\right]
\end{eqnarray}
for $B>0$, $C>0$, and $L>0$.
The covariance function in \eq{rot-kernel} cannot exactly reproduce \eq{sine2}
but, since \eq{sine2} is only an effective model, \eq{rot-kernel} can be used
as a drop-in replacement for a significant gain in computational efficiency.

\response{GPs have been used to measure stellar rotation periods for
individual datasets \citep[for example][]{Littlefair:2017}, but the
computational cost of traditional GP methods has hindered the industrial
application to existing surveys like \kepler\ with hundreds of thousands of
targets.
The increase in computational efficiency and scalability provided by
\celerite\ opens the possibility of inferring rotation periods using GPs at
scale of existing and forthcoming surveys like \kepler, \tess, and \lsst.
}

As a demonstration, we fit a \celeriteterm\ model with a kernel given
by \eq{rot-kernel} to a \kepler\ light curve for the star KIC~1430163.
This star has a published rotation period of $3.88 \pm 0.58\,\mathrm{days}$,
measured using traditional periodogram and autocorrelation function approaches
applied to \kepler\ data from Quarters 0--16 \citep{Mathur:2014}, covering
about four years.

\response{We select about 180~days of contiguous observations of
KIC~1430163 from \kepler.
This dataset has \exampleiiindata\ measurements and, using a tuned linear
algebra
implementation\footnote{We use the Intel Math Kernel Library
\url{https://software.intel.com/en-us/intel-mkl}}, a
single evaluation of the likelihood requires over 8~seconds on a modern Intel
CPU.
This calculation, using \celerite\ with the model in \eq{rot-kernel}, only
takes $\sim 1.5\,\mathrm{ms}$~--~a speed-up of more than three orders of
magnitude per model evaluation. 

We set the mean function $\mu_\bvec{\theta}$ to zero and the remaining
parameters $\bvec{\alpha}$ and their priors are listed in
Table~\ref{tab:example-3-params}.
As with the earlier examples, we start by estimating the MAP parameters using
\project{L-BFGS-B} and initialize \exampleiiinwalkers~walkers by sampling from
an isotropic Gaussian with a standard deviation of $10^{-5}$ centered on the
MAP parameters.
The left panels of \Figure{rotation} show a subset of the data used in this
example and the residuals away from the MAP predictive mean.

We run \exampleiiinburn~steps of burn-in, followed by \exampleiiinsteps~steps
of MCMC using \emcee.
We estimate the integrated autocorrelation time of the chain for $\ln
P_\mathrm{rot}$ and estimate that we have \exampleiiineff\ independent
samples.
These samples give a posterior constraint on the period of $P_\mathrm{rot} =
\rotationperiod$ days and the marginalized posterior distribution for $P$ is
shown in the right panel of \Figure{rotation}.
This result is in good agreement with the literature value with smaller
uncertainties.
A detailed comparison of GP rotation period measurements and the traditional
methods is beyond the scope of this paper, but \citet{Angus:2017} demonstrate
that GP inferences are, at a population level, more reliable than other
methods.

The total computational cost for this inference using \celerite\ is about
4~CPU-minutes.
By contrast, the same inference using a general but optimized Cholesky
factorization routine would require nearly 400~CPU-hours.
This speed-up enables probabilistic measurement of rotation periods using
existing data from \project{K2} and forthcoming surveys like \tess\ and \lsst\
where this inference will need to be run for at least hundreds of thousands of
stars.
}

\begin{floattable}
\begin{deluxetable}{cc}
\tablecaption{The parameters and priors for Example 3. \label{tab:example-3-params}}
\tablehead{\colhead{parameter} & \colhead{prior}}
\startdata
$\ln(B/\mathrm{ppt}^2)$ & $\mathcal{U}(-10.0,\,0.0)$ \\
$\ln(L/\mathrm{day})$ & $\mathcal{U}(1.5,\,5.0)$ \\
$\ln(P/\mathrm{day})$ & $\mathcal{U}(-3.0,\,5.0)$ \\
$\ln(C)$ & $\mathcal{U}(-5.0,\,5.0)$ \\
\enddata
\end{deluxetable}
\end{floattable}

\begin{figure}[hptb]
\begin{center}
\includegraphics[width=0.9\textwidth]{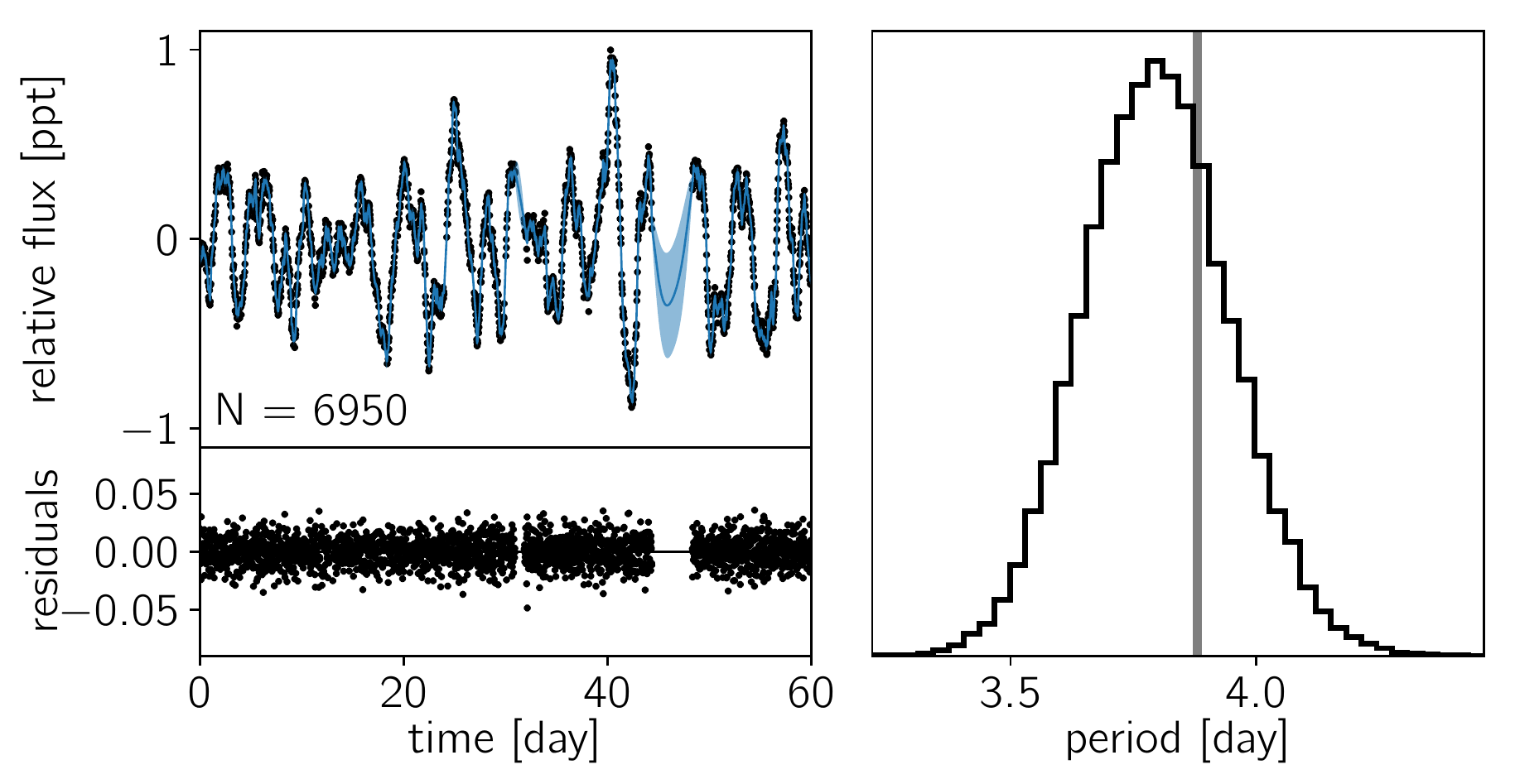}
\caption{Inferred constraints on a quasiperiodic GP model using the covariance
    function in \eq{rot-kernel} and two quarters of \kepler\ data.
(top left) The \kepler\ data (black points) and the MAP model
    prediction (blue curve) for a 60~day subset of the data used.
    The solid blue line shows the predictive mean and the blue contours show
    the predictive standard deviation.
(bottom left) The residuals between the mean predictive model and the data
    shown in the top left figure.
(right) The posterior constraint on the rotation period of KIC~1430163 using
    the dataset and model from \Figure{rotation}.
    The period is the parameter $P_\mathrm{rot}$ in \eq{rot-kernel} and this
    figure shows the posterior distribution marginalized over all other
    nuisance parameters in \eq{rot-kernel}.  This is consistent with the
    published rotation period made using the full
    \kepler\ baseline shown as a vertical gray line \citep{Mathur:2014}.
    \figurelabel{rotation}}
\end{center}
\end{figure}

\subsection{Example 4: Asteroseismic oscillations}\sectlabel{astero}

The asteroseismic oscillations of thousands of stars were measured using light
curves from the \kepler\ Mission \citep{Gilliland:2010, Huber:2011,
Chaplin:2011, Chaplin:2013, Stello:2013} and asteroseismology is a key science
driver for many of the upcoming large scale photometric surveys
\citep{Campante:2016, Rauer:2014, Gould:2015}.
Most asteroseismic analyses have been limited to high signal-to-noise
oscillations because the standard methods use statistics of the empirical
periodogram.
These methods cannot formally propagate the measurement uncertainties to the
constraints on physical parameters and they instead rely on population-level
bootstrap uncertainty estimates \citep{Huber:2009}.
More sophisticated methods that compute the likelihood function in the time
domain scale poorly to large survey datasets \citep{Brewer:2009,
Corsaro:2014}.

\celerite\ alleviates these problems by providing a physically motivated
probabilistic model that can be evaluated efficiently even for large datasets.
In practice, we model the star as a mixture of stochastically-driven simple
harmonic oscillators where the amplitudes and frequencies of the oscillations
are computed using a physical model, and evaluate the probability of the
observed time series using a GP where the PSD is a sum of terms given by
\eq{sho-psd}.
This gives us a method for computing the likelihood function for the
parameters of the physical model (for example, $\nu_\mathrm{max}$ and $\Delta
\nu$, or other more fundamental parameters) \emph{conditioned on the observed
time series} in $\mathcal{O}(N)$ operations.
In other words, \celerite\ provides a computationally efficient framework that
can be combined with physically-motivated models of stars and numerical
inference methods to make rigorous probabilistic measurements of asteroseismic
parameters in the time domain.
This has the potential to push asteroseismic analysis to lower signal-to-noise
datasets and we hope to revisit this idea in a future paper.

To demonstrate the method, we use a simple heuristic model where the PSD is
given by a mixture of 8 components with amplitudes and frequencies specified
by $\nu_\mathrm{max}$, $\Delta \nu$, and some nuisance parameters.
The first term is used to capture the granulation ``background''
\citep{Kallinger:2014} using \eq{granulation-psd} with two free parameters
$S_g$ and $\omega_g$.
The remaining  7 terms are given by \eq{sho-psd} where $Q$ is a nuisance
parameter shared between terms and the frequencies are given by
\begin{eqnarray}
\omega_{0,\,j} = 2\,\pi\,(\nu_\mathrm{max} + j\,\Delta\nu + \epsilon)
\end{eqnarray}
and the amplitudes are given by
\begin{eqnarray}
S_{0,\,j} =
    \frac{A}{Q^2}\,\exp\left(-\frac{[j\,\Delta\nu + \epsilon]^2}{2\,W^2}\right)
\end{eqnarray}
where $j$ is an integer running from $-3$ to 3 and $\epsilon$, $A$, and $W$
are shared nuisance parameters.
\response{Finally, we also fit for the amplitude of the white noise by adding
a parameter $\sigma$ in quadrature with the uncertainties given for the
\kepler\ observations.
All of these parameters and their chosen priors are listed in
Table~\ref{tab:example-4-params}.
As before, these priors are all log-uniform except for $\epsilon$ where we use
a zero-mean normal prior with a broad variance of $1~\mathrm{day}^2$ to break
the degeneracy between $\nu_\mathrm{max}$ and $\epsilon$.
}
To build a more realistic model, this prescription could be extended to
include more angular modes, or $\nu_\mathrm{max}$ and $\Delta \nu$ could be
replaced by the fundamental physical parameters of the star.

To demonstrate the applicability of this model, we apply it to infer the
asteroseismic parameters of the giant star KIC~11615890, observed by the
\kepler\ Mission.
The goal of this example is to show that, even for a low signal-to-noise
dataset with a short baseline, it is possible to infer asteroseismic
parameters with formal uncertainties that are consistent with the parameters
inferred with a much larger dataset.
Looking forward to \tess\ \citep{Ricker:2014,Campante:2016}, we measure
$\nu_\mathrm{max}$ and $\Delta\nu$ using only one month of \kepler\ data and
compare our results to the results inferred from the full 4 year baseline of
the \kepler\ Mission.
For KIC~11615890, the published asteroseismic parameters measured using several
years of \kepler\ observations are \citep{Pinsonneault:2014}
\begin{eqnarray}
    \nu_\mathrm{max} = 171.94 \pm 3.62 \,\mu\mathrm{Hz} \quad\mathrm{and}\quad
    \Delta\nu = 13.28 \pm 0.29 \,\mu\mathrm{Hz} \quad.
\end{eqnarray}

Unlike typical giants, KIC~11615890 is a member of a class of stars where the
dipole ($\ell = 1$) oscillation modes are suppressed by strong magnetic fields
in the core \citep{Stello:2016}.
This makes this target simpler for the purposes of this demonstration because
we can neglect the $\ell = 1$ modes and the model proposed above will be an
effective model for the combined signal from the $\ell = 0$ and 2 modes.

\response{For this demonstration, we randomly select a month-long segment of
PDC \kepler\ data \citep{Stumpe:2012, Smith:2012}.
Unlike the previous examples, the posterior distribution is sharply
multimodal and na\"ively maximizing the posterior using \project{L-BFGS-B} is
not practical.
Instead, we start by estimating the initial values for $\nu_\mathrm{max}$ and
$\Delta\nu$ using only the month-long subset of data and following the
standard procedure described by \citet{Huber:2009}.
We then run a set of \project{L-BFGS-B} optimizations with values of
$\nu_\mathrm{max}$ selected in logarithmic grid centered on our initial
estimate of $\nu_\mathrm{max}$ and initial values of $\Delta\nu$ computed
using the empirical relationship between these two quantities
\citep{Stello:2009}.
This initialization procedure is sufficient for this example, but the general
application of this method will require a more sophisticated prescription.

\Figure{astero} shows a 10~day subset of the dataset used for this example.
The MAP model is overplotted on these data and the residuals away from the
mean prediction of this model are shown in the bottom panel of
\Figure{astero}.
There is no obvious structure in these residuals, lending some credibility to
the model specification.

We initialize \exampleivnwalkers~walkers by sampling from an isotropic
Gaussian centered on the MAP parameters (the full set of parameters and their
priors are listed in Table~\ref{tab:example-4-params}), run
\exampleivnburn~steps of burn-in, and run \exampleivnsteps~steps of MCMC using
\emcee.
We estimate the mean autocorrelation time for the chains of
$\ln\nu_\mathrm{max}$ and $\ln\Delta\nu$ and find \exampleivneff\ effective
samples from the marginalized posterior density.
\Figure{astero-corner} shows the marginalized density for $\nu_\mathrm{max}$
and $\Delta\nu$ compared to the results from the literature.
This result is consistent within the published error bars and the posterior
constraints are tighter than the published results.
The top two panels of \Figure{astero_psd} show the Lomb-Scargle periodogram
\citep{VanderPlas:2017} estimator for the power spectrum based on the full
4~years of \kepler\ data and the month-long subset used in our analysis.
The bottom panel of \Figure{astero_psd} shows the posterior inference of the
PSD using the \celerite\ and the model described here.
Despite only using one month of data, the \celerite\ inference captures the
salient features of the power spectrum and it is qualitatively consistent with
the standard estimator applied to the full baseline.
All asteroseismic analyses are known to have substantial systematic and
method-dependent uncertainties \citep{Verner:2011} so further experiments
would be needed to fully assess the reliability of this specific model.

This model requires about 10~CPU minutes to run the MCMC to convergence.
This is more computationally intensive than traditional methods of measuring
asteroseismic oscillations, but it is much cheaper than the same analysis
using a general direct GP solver.
For comparison, we estimate that repeating this analysis using a general
Cholesky factorization implemented as part of a tuned linear algebra
library\footnote{We use the Intel Math Kernel Library
\url{https://software.intel.com/en-us/intel-mkl}} would require about 15~CPU
hours.
An in-depth discussion of the benefits of rigorous probabilistic inference of
asteroseismic parameters in the time domain is beyond the scope of this paper,
but we hope to revisit this opportunity in the future.
}

\begin{floattable}
\begin{deluxetable}{cc}
\tablecaption{The parameters and priors for Example 4. \label{tab:example-4-params}}
\tablehead{\colhead{parameter} & \colhead{prior}}
\startdata
$\ln(S_g/\mathrm{ppm}^2)$ & $\mathcal{U}(-15,\,15)$ \\
$\ln(\omega_g/\mathrm{day}^{-1})$ & $\mathcal{U}(-15,\,15)$ \\
$\ln(\nu_\mathrm{max}/\mu\mathrm{Hz})$ & $\mathcal{U}(\ln(130),\,\ln(190))$ \\
$\ln(\Delta \nu/\mu\mathrm{Hz})$ & $\mathcal{U}(\ln(12.5),\,\ln(13.5))$ \\
$\epsilon/\mathrm{day}^{-1}$ & $\mathcal{N}(0,\,1)$ \\
$\ln(A/\mathrm{ppm}^2\,\mathrm{day})$ & $\mathcal{U}(-15,\,15)$ \\
$\ln(Q)$ & $\mathcal{U}(-15,\,15)$ \\
$\ln(W/\mathrm{day}^{-1})$ & $\mathcal{U}(-3,\,3)$ \\
$\ln(\sigma/\mathrm{ppm})$ & $\mathcal{U}(-15,\,15)$ \\
\enddata
\end{deluxetable}
\end{floattable}

\begin{figure}[!hptb]
\begin{center}
\includegraphics[width=0.5\textwidth]{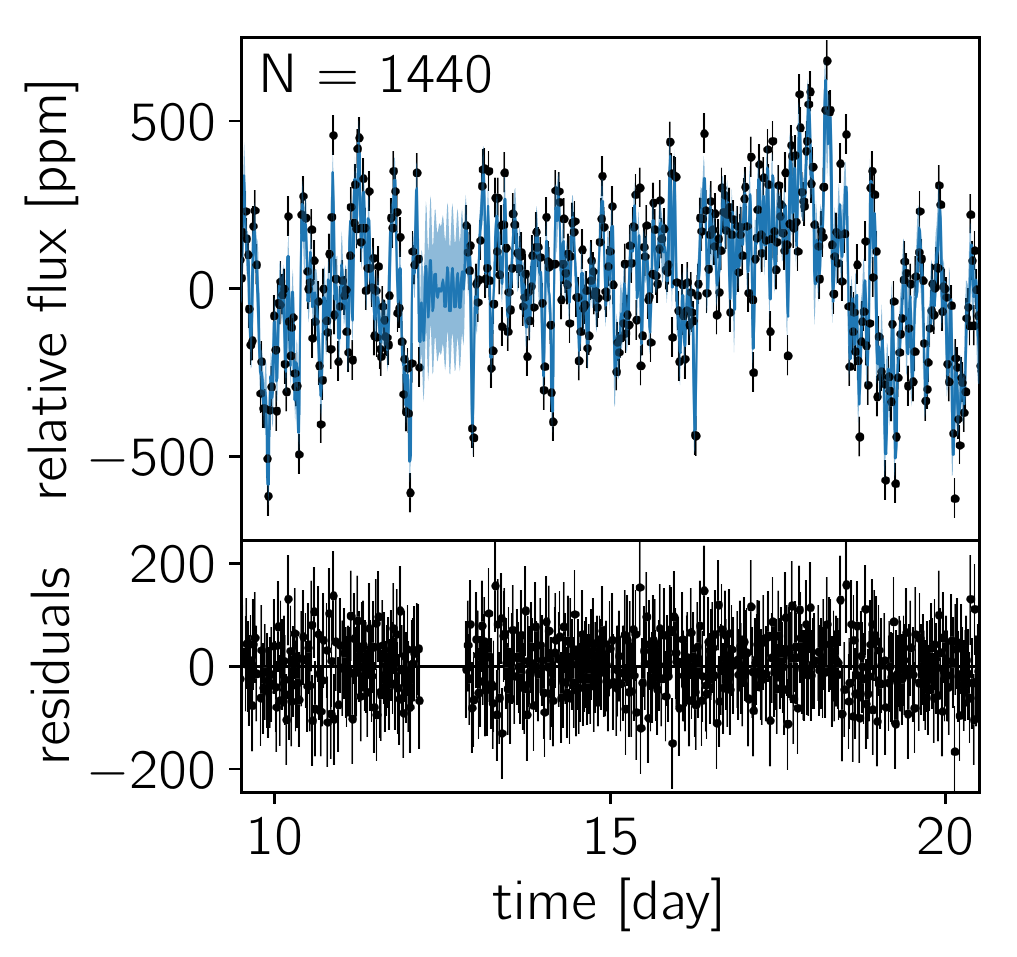}
\caption{
\response{
(top) The \kepler\ data (black points) and the MAP model
    prediction (blue curve) for a 10~day subset of the month-long dataset that
    was used for the fit.
    The solid blue line shows the predictive mean and the blue contours show
    the predictive standard deviation.
(bottom) The residuals between the mean predictive model and the data
    shown in the top figure.
}
    \figurelabel{astero}}
\end{center}
\end{figure}

\begin{figure}[!p]
\begin{center}
\includegraphics[width=0.8\textwidth]{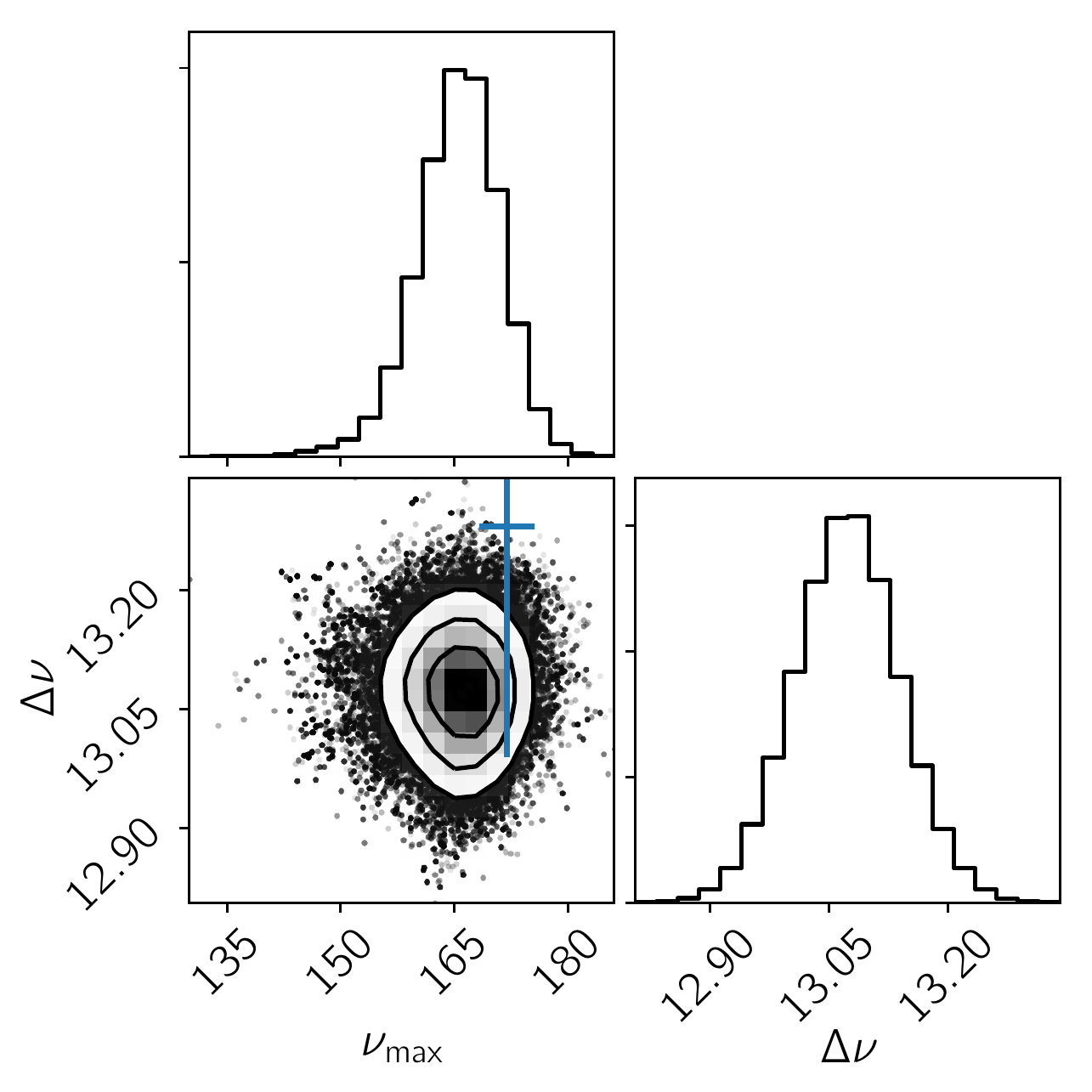}
\caption{The probabilistic constraints on $\nu_\mathrm{max}$ and $\Delta \nu$
    from the inference shown in \Figure{astero} compared to the published
    value (error bar) based on several years of \kepler\ observations
    \citep{Pinsonneault:2014}.
    The two-dimensional contours show the 0.5-, 1-, 1.5, and 2-sigma credible
    regions in the marginalized planes and the histograms along the diagonal
    show the marginalized posterior for each parameter.
    \figurelabel{astero-corner}}
\end{center}
\end{figure}

\begin{figure}[!p]
\begin{center}
\includegraphics[width=0.9\textwidth]{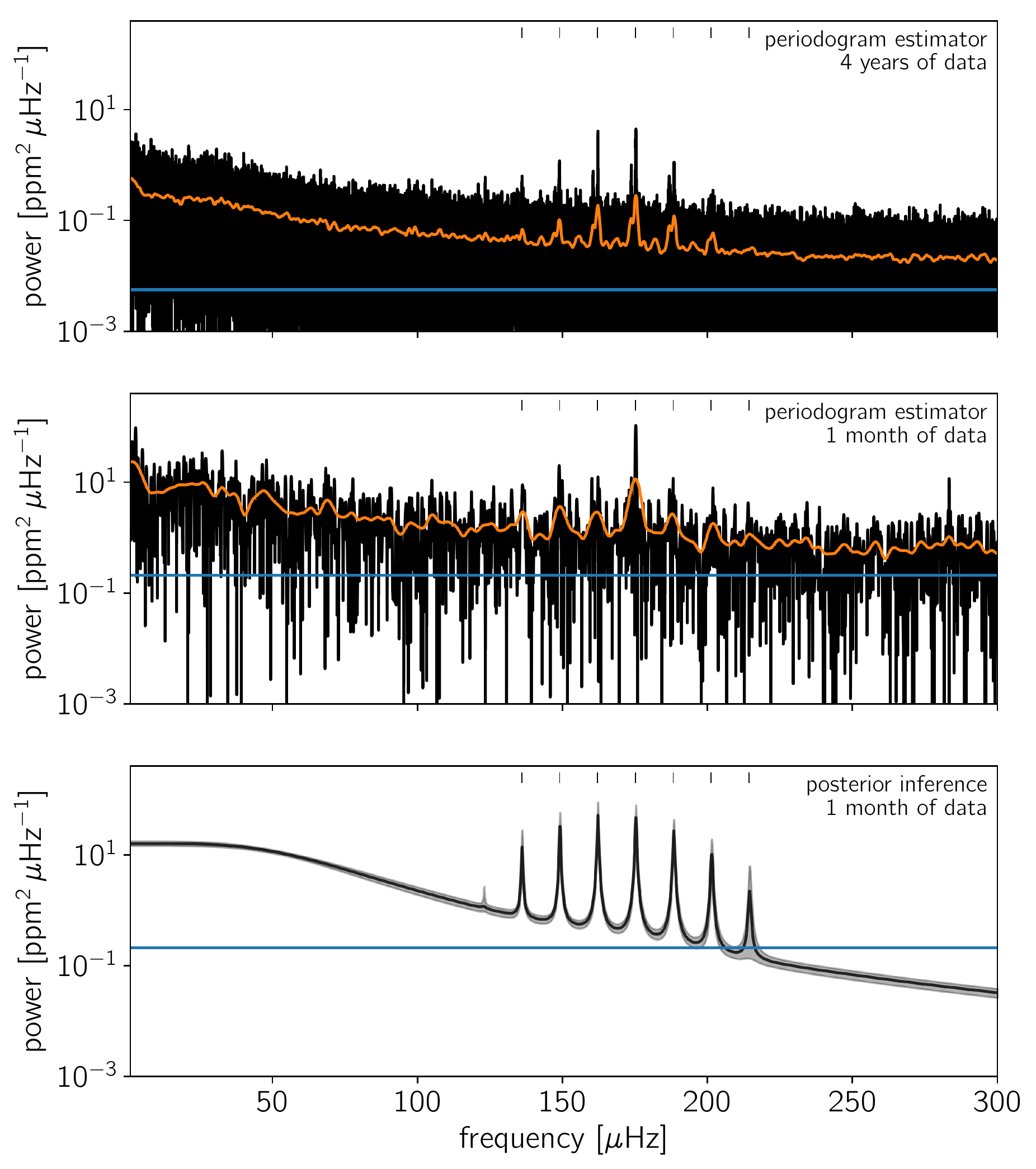}
\caption{A comparison between the Lomb-Scargle estimator of the PSD and the
posterior inference of the PSD as a mixture of stochastically-driven simple
harmonic oscillators.
(top) The periodogram of the \kepler\ light curve for KIC~11615890 computed
    on the full four year baseline of the mission. The orange line shows a
    smoothed periodogram and the blue line indicates the level of the
    measurement uncertainties.
(middle) The same periodogram computed using about a month of data.
(bottom) The power spectrum inferred using the mixture of SHOs model described
    in the text and only one month of \kepler\ data.
    The black line shows the median of posterior PSD and the gray contours
    show the 68\% credible region.
    \figurelabel{astero_psd}}
\end{center}
\end{figure}

\newpage
\subsection{Example 5: Exoplanet transit fitting}\sectlabel{transit}

\response{
This example is different from all the previous examples~--~both simulated and
real~--~because, in this case, do not set the deterministic
mean function $\mu_\bvec{\theta}$ to zero.
Instead, we make inferences about $\mu_\bvec{\theta}$ because it is a
physically significant function and the parameters are fundamental properties
of the system.
This is an example using real data~--~because we use the light curve from
\sect{rotation}~--~but we multiply these data by a simulated transiting
exoplanet model with known parameters.
This allows us to show that we can recover the true parameters of the planet
even when the transit signal is superimposed on the real variability of a
star.
}
GP modeling has been used extensively for this purpose throughout the
exoplanet literature \citep[for example][]{Dawson:2014, Barclay:2015,
Evans:2015, Foreman-Mackey:2016b, Grunblatt:2016}.

In \eq{gp-likelihood} the physical parameters of the exoplanet are called
$\bvec{\theta}$ and, in this example, the mean function $\mu_\bvec{\theta}(t)$
is a limb-darkened transit light curve \citep{Mandel:2002,
Foreman-Mackey:2016a} and the parameters $\bvec{\theta}$ are the orbital
period $P_\mathrm{orb}$, the transit duration $T$, the phase or epoch $t_0$,
the impact parameter $b$, the radius of the planet in units of the stellar
radius $R_P/R_\star$, the baseline relative flux of the light curve $f_0$, and
two parameters describing the limb-darkening profile of the star
\citep{Claret:2011, Kipping:2013}.
As in \sect{rotation}, we model the stellar variability using a GP model
with a kernel given by \eq{rot-kernel} and fit for the parameters of the
exoplanet $\bvec{\theta}$ and the stellar variability $\bvec{\alpha}$
simultaneously.
\response{The full set of parameters $\bvec{\alpha}$ and $\bvec{\theta}$ are
listed in Table~\ref{tab:example-5-params} along with their priors and the
true values for $\bvec{\theta}$.}

\response{ We take a 20~day segment of the \kepler\ light curve for
KIC~1430163 ($N = \examplevndata$) and multiply it by a simulated transit
model with the parameters listed in Table~\ref{tab:example-5-params}.
Using these data, we maximize the joint posterior defined by the likelihood in
\eq{gp-likelihood} and the priors in Table~\ref{tab:example-5-params} using
\project{L-BFGS-B} for all the parameters $\bvec{\alpha}$ and $\bvec{\theta}$
simultaneously.
The top panel of \Figure{transit-ml} shows the data including the simulated
transit as black points with the MAP model prediction over-plotted in blue.
The bottom panel of \Figure{transit-ml} shows the ``de-trended'' light curve
where the MAP model has been subtracted and the MAP mean model
$\mu_\bvec{\theta}$ has been added back in.
For comparison, the transit model is over-plotted in the bottom panel of
\Figure{transit-ml} and we see no obvious correlations in the residuals.

Sampling \examplevnwalkers~walkers from an isotropic Gaussian centered on the
MAP parameters, we run \examplevnburn~steps of burn-in and
\examplevnsteps~steps of production MCMC using \emcee.
We estimate the integrated autocorrelation time for the $\ln(P_\mathrm{orb})$
chain and find \examplevneff~effective samples across the full chain.
\Figure{transit-corner} shows the marginalized posterior constraints on the
physical properties of the planet compared to the true values.
Even though the \celeriteterm\ representation of the stellar variability is
only an effective model, the inferred distributions for the physical
properties of the planet $\bvec{\theta}$ are consistent with the true values.
This promising result suggests that \celerite\ can be used as an effective
model for transit inference for a significant gain in computational
tractability.
Even for this example with small $N$, the cost of computing the likelihood
using \celerite\ is nearly two orders of magnitude faster than the same
computation using a general direct solver.
}

\begin{floattable}
\begin{deluxetable}{lccc}
\tablecaption{The parameters, priors, and (if known) the true values used for
the simulation in Example 5. \label{tab:example-5-params}}
\tablehead{& \colhead{parameter} & \colhead{prior} & \colhead{true value}}
\startdata
\multirow{5}{*}{kernel: $\bvec{\alpha} \quad$}
& $\ln(B/\mathrm{ppt}^2)$ & $\mathcal{U}(-10.0,\,0.0)$ & --- \\
& $\ln(L/\mathrm{day})$ & $\mathcal{U}(1.5,\,5.0)$ & --- \\
& $\ln(P_\mathrm{rot}/\mathrm{day})$ & $\mathcal{U}(-3.0,\,5.0)$ & --- \\
& $\ln(C)$ & $\mathcal{U}(-5.0,\,5.0)$ & --- \\
& $\ln(\sigma/\mathrm{ppt})$ & $\mathcal{U}(-5.0,\,0.0)$ & --- \\
\hline
\multirow{8}{*}{mean: $\bvec{\theta} \quad$}
& $f_0/\mathrm{ppt}$ & $\mathcal{U}(-0.5,\,0.5)$ & 0 \\
& $\ln(P_\mathrm{orb}/\mathrm{day})$ & $\mathcal{U}(\ln(7.9),\,\ln(8.1))$ & $\ln(8)$  \\
& $\ln(R_p/R_\star)$ & $\mathcal{U}(\ln(0.005),\,\ln(0.1))$ & $\ln(0.015)$ \\
& $\ln(T/\mathrm{day})$ & $\mathcal{U}(\ln(0.4),\,\ln(0.6))$ & $\ln(0.5)$ \\
& $t_0/\mathrm{day}$ & $\mathcal{U}(-0.1,\,0.1)$ & $0$ \\
& $b$ & $\mathcal{U}(0,\,1.0)$ & $0.5$ \\
& $q_1$ & $\mathcal{U}(0,\,1)$ & $0.5$ \\
& $q_2$ & $\mathcal{U}(0,\,1)$ & $0.5$ \\
\enddata
\end{deluxetable}
\end{floattable}

\begin{figure}[!hptb]
\begin{center}
\includegraphics[width=\textwidth]{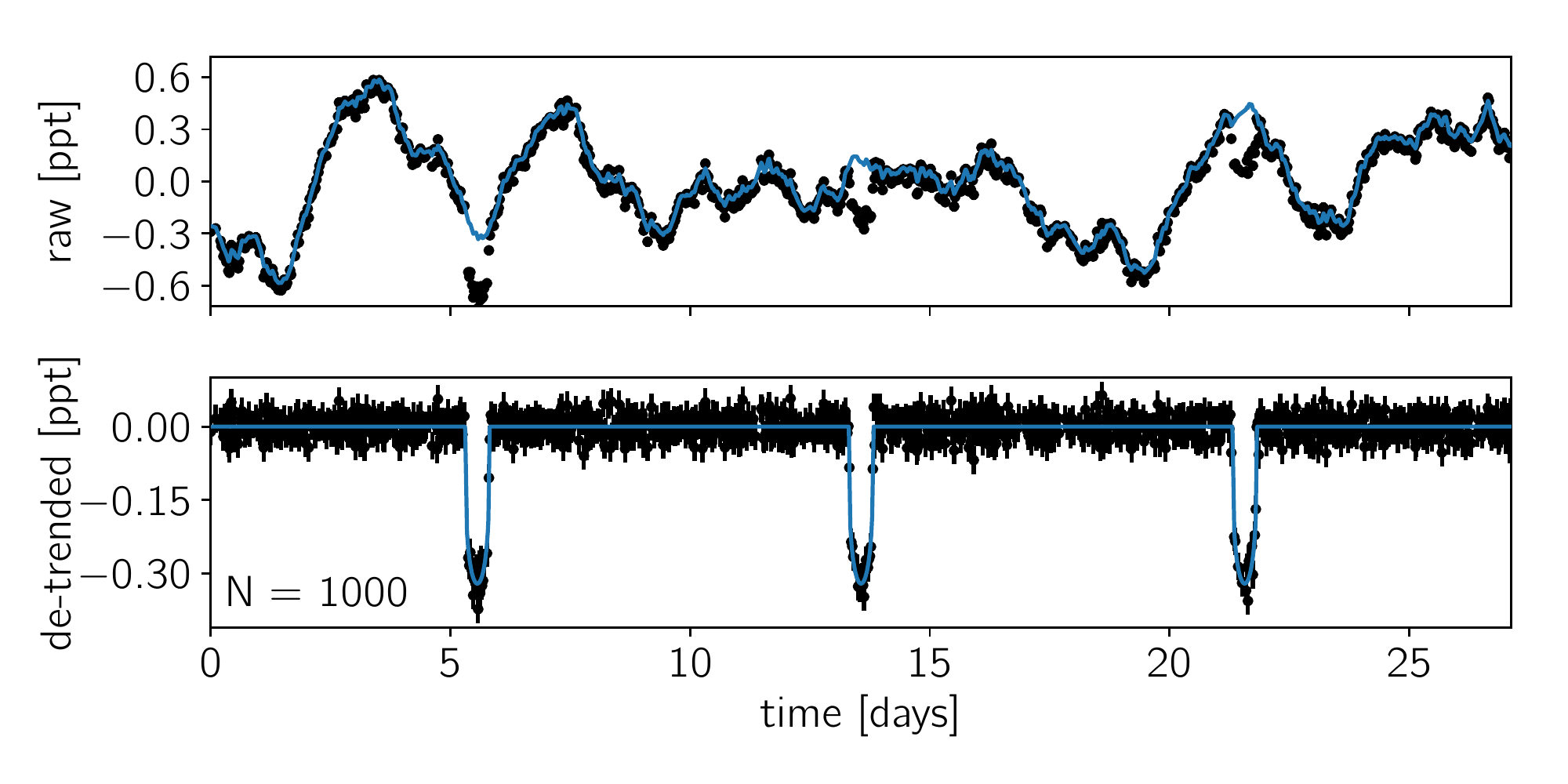}
    \caption{\emph{(top)} A month-long segment of \kepler\ light curve for
    KIC~1430163 with a synthetic transit model injected (black points) and the
    MAP model for the stellar variability
    (blue line).
    \emph{(bottom)} The MAP ``de-trending'' of the data in
    the top panel.
    In this panel, the MAP model for the stellar variability
    has been subtracted to leave only the transits.
    The de-trended light curve is shown by black error bars and the MAP
    transit model is shown as a blue line.
    \figurelabel{transit-ml}}
\end{center}
\end{figure}

\begin{figure}[!p]
\begin{center}
\includegraphics[width=\textwidth]{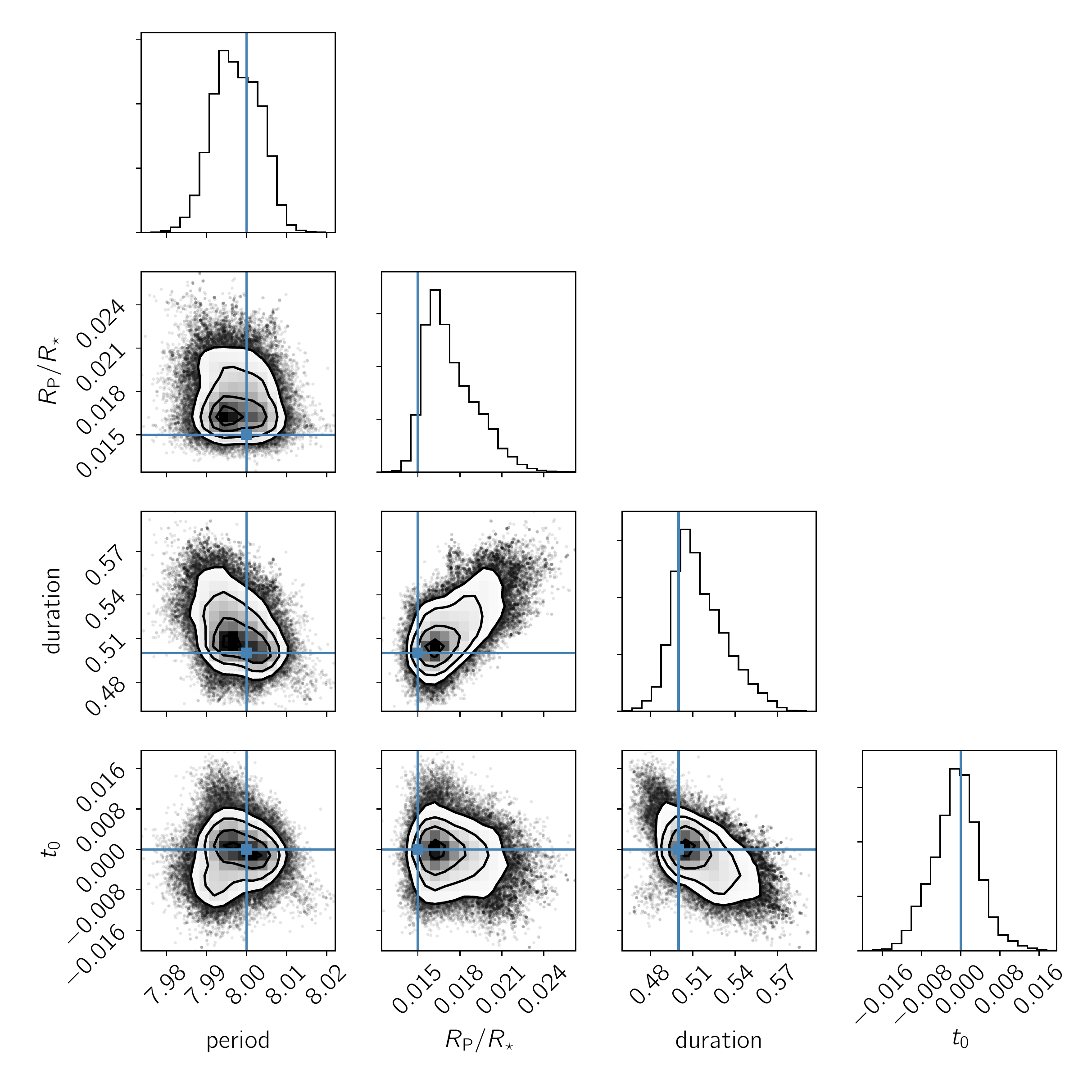}
\caption{The marginalized posterior constraints on the physical parameters of
    the planet transit in the light curve shown in the top panel of
    \Figure{transit-ml}.
    The two-dimensional contours show the 0.5-, 1-, 1.5, and 2-sigma credible
    regions in the marginalized planes and the histograms along the diagonal
    show the marginalized posterior for each parameter.
    The true values used in the simulation are indicated by blue lines.
    For each parameter, the inference is consistent with the true value.
    \figurelabel{transit-corner}}
\end{center}
\end{figure}

\subsection{Summary}

\response{In the previous subsections, we demonstrate five potential use
cases for \celerite.
These examples span a range of data sizes and model complexities, and the last
three are based on active areas of research with immediate real-world
applicability.
The sizes of these datasets are modest on the scale of some existing and
forthcoming survey datasets because we designed the experiments to be easily
reproducible but, even in these cases, the inferences made here would be
intractable without substantial computational investment.

Table~\ref{tab:example-stats} lists the specifications of each example.
In this table, we list the computational cost required for a single likelihood
evaluation using a general Cholesky factorization and the cost of computing
the same likelihood using \celerite.
Combining this with the total number of model evaluations required to run the
MCMC algorithm to convergence (this number is also listed in
Table~\ref{tab:example-stats}), we can estimate the total CPU time required in
each case.
While the exact computational requirements for any problem also depend on the
choice of inference algorithm and the specific precision requirements, this
table provides a rough estimate of what to expect for projects at these
scales.
}

\begin{floattable}
\begin{deluxetable}{cccccccc}
\caption{The computational cost and convergence stats for each example. \label{tab:example-stats}}
\tablehead{& \colhead{$N$} & \colhead{$J$} & \colhead{direct\tablenotemark{a}} & \colhead{\celerite\tablenotemark{b}} & \colhead{dimension\tablenotemark{c}} & \colhead{evaluations\tablenotemark{d}} & \colhead{$N_\mathrm{eff}$\tablenotemark{e}}\\
&&& \colhead{ms} & \colhead{ms} &&&}
\startdata
1 & 200 & 1 & 2.85 & 0.26 & 3 & 80000 & 1737 \\
2 & 100 & 2 & 0.67 & 0.36 & 4 & 80000 & 1134 \\
3 & 6950 & 2 & 8119.11 & 1.47 & 4 & 176000 & 2900 \\
4 & 1440 & 8 & 828.93 & 2.74 & 9 & 640000 & 1443 \\
5 & 1000 & 2 & 88.30 & 0.96 & 13 & 1280000 & 1490 \\
\enddata

\tablenotetext{a}{The computational cost of computing the GP model using the
general Cholesky factorization routine implemented in the Intel MKL.}
\tablenotetext{b}{The computational cost of computing the GP model using
\celerite.}
\tablenotetext{c}{The total number of parameters in the model.}
\tablenotetext{d}{The total number of evaluations of the model to run MCMC to
convergence.}
\tablenotetext{e}{The effective number of independent samples estimated by
computing the integrated autocorrelation time of the chain.}
\end{deluxetable}
\end{floattable}

\section{Comparisons to other methods}\sectlabel{compare}

There are other methods of scaling GP models to large datasets and in this
section we draw comparisons between \celerite\ and other popular methods.
Scalable GP methods tend to fall into two categories: approximate and
restrictive.
\celerite\ falls into the latter category because, while the method is exact,
it requires a specific choice of stationary kernel function and it can only be
used in one-dimension.

The continuous autoregressive moving average (CARMA) models introduced to the
astrophysics literature by \citet{Kelly:2014} share many features with
\celerite.
CARMA models are restricted to one-dimensional problems and the likelihood
function for a CARMA model can be solved in $\mathcal{O}(N\,J^2)$ using a
Kalman filter \citep{Kelly:2014}.
The kernel function for a CARMA$(J,\,K)$ model is
\begin{eqnarray}\eqlabel{carma-kernel}
k_\mathrm{CARMA}(\tau) = \sum_{j=1}^J A_j\,\exp\,(r_j\,\tau)
\end{eqnarray}
where
\begin{eqnarray}\eqlabel{carma-coeff}
A_j = \sigma^2 \,\frac{\left[\sum_{k=0}^K\beta_k\,{(r_j)}^k\right]\,
    \left[\sum_{k=0}^K\beta_k\,{(-r_j)}^k\right]}
    {-2\,\mathrm{Re}(r_j)\,\prod_{k=1,\,k \ne j}^{J}(r_k-r_j)({r_k}^*+r_j)}
\end{eqnarray}
and $\sigma$, $\{r_j\}_{j=1}^J$ and $\{\beta_k\}_{k=1}^K$ are parameters of
the model; $\sigma$ is real, while the others are complex.
Comparing \eq{carma-kernel} to \eq{celerite-kernel-complex}, we can see that
every CARMA model corresponds to an equivalent \celeriteterm\ model and the
parameters $a_j$, $b_j$, $c_j$, and $d_j$ can be easily computed analytically.
The inverse statement is not as simple.
In practice, this means that \celerite\ could be trivially used to compute any
CARMA model.
Using the CARMA solver to compute a \celeriteterm\ model, however, requires
solving \eq{carma-coeff} numerically for a given set of $\{A_j\}_{j=1}^J$.

\response{
The computational scaling of CARMA models is also $\mathcal{O}(N\,J^2)$ using
the Kalman filter method \citep{Kelly:2014}, but we find that, in practice,
this method is more computationally expensive than \celerite, but it has a
smaller memory footprint.
\Figure{benchmark-carma} shows the scaling of the Kalman filter solver for the
systems shown in \Figure{benchmark}.
Comparing these figures, we find that the Kalman filter solver is slower than
\celerite\ for all the systems we tested with an average difference of about
an order of magnitude.

Another benefit of \celerite\ compared to Kalman filter solvers is the fact
that it is an algebraic solver for the relevant matrix equations instead of
an algorithm to directly compute \eq{gp-likelihood}.
This means that the inverse covariance matrix $K^{-1}$ can be applied to
general vectors and matrices in $\mathcal{O}(N)$, while reusing the
factorization.
This is a crucial feature when GP models are combined with linear regression
\citep[for example][]{Luger:2017}.
}

\begin{figure}[tp]
\begin{center}
\includegraphics[width=\textwidth]{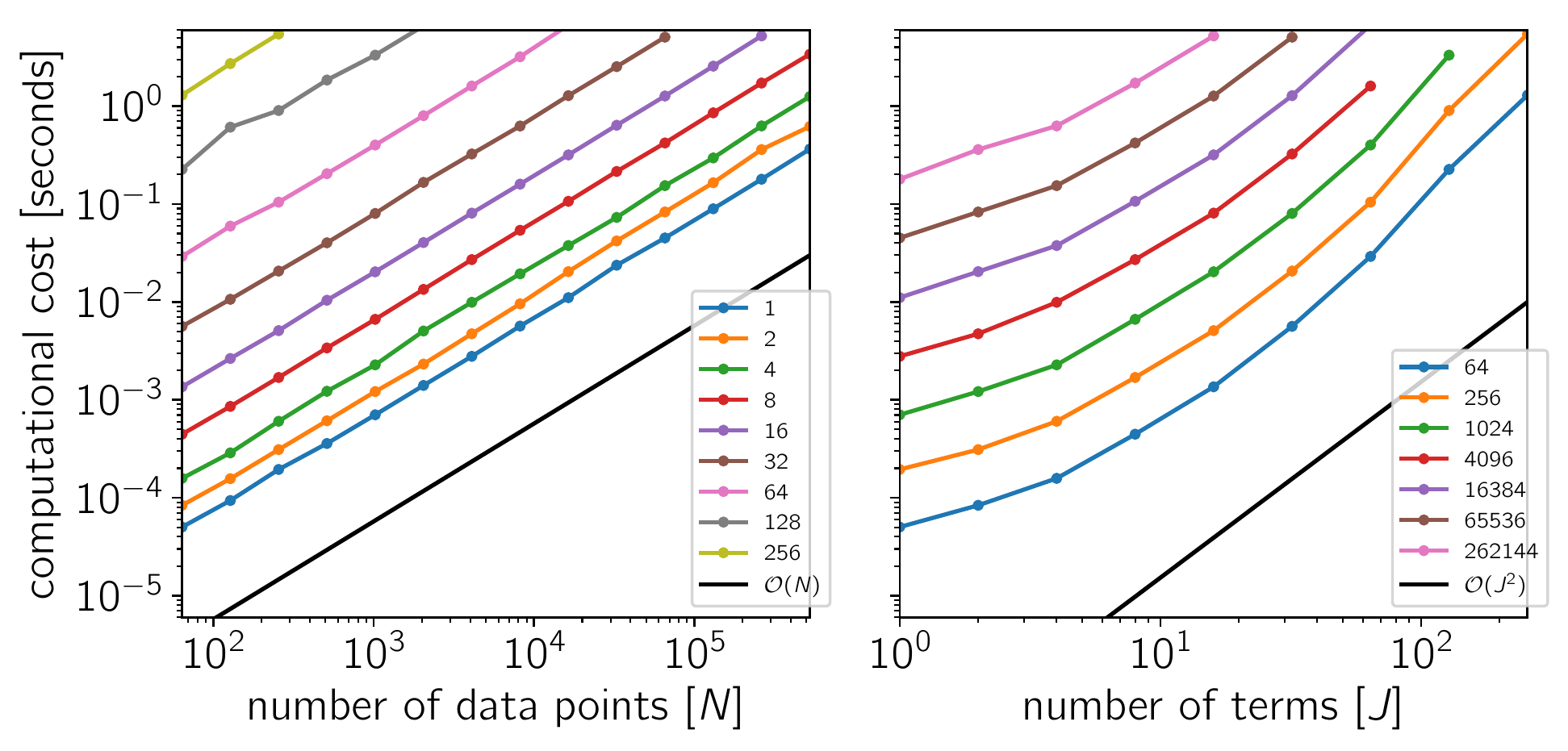}
\caption{\response{The same as \Figure{benchmark}, but using an optimized
    \project{C++} implementation of the Kalman filter method developed by
    \citet{Kelly:2014}.
    Comparing this figure to \Figure{benchmark}, we see that \celerite\ is
    about an order of magnitude faster in all cases.
    \emph{(left)} The cost of computing \eq{gp-likelihood} with a covariance
    matrix given by \eq{celerite-kernel} as a function of the number of data
    points $N$.
    The different lines show the cost for different numbers of terms $J$
    increasing from bottom to top.
    To guide the eye, the straight black line without points
    shows linear scaling in $N$.
    \emph{(right)} The same information plotted as a function of $J$ for
    different values of $N$.
    Each line shows the scaling for a specific value of $N$ increasing from
    bottom to top.
    The black line shows quadratic scaling in $J$.}
    \figurelabel{benchmark-carma}}
\end{center}
\end{figure}

Another popular GP method uses the fact that, in the limit of evenly-spaced
data and homoscedastic uncertainties, the covariance matrix is ``Toeplitz''
\citep[for example][]{Dillon:2013}.
There are exact methods for solving Toeplitz matrix equations that scale as
$\mathcal{O}(N\,\log N)$ and methods for computing determinants exactly in
$\mathcal{O}(N^2)$ or approximately in $\mathcal{O}(N\,\log N)$
\citep{Wilson:2014}.
The Toeplitz method is, in some ways, more flexible than \celerite\ because it
can be used with any stationary kernel, but it requires uniformly spaced data
and the scaling is worse than \celerite\ so it is less efficient when applied
to large datasets.

\citet{Carter:2009} improved the scaling of Toeplitz methods by introducing a
wavelet-based method for computing a GP likelihood with $\mathcal{O}(N)$
scaling.
This method has been widely applied in the context of exoplanet transit
characterization, but it requires evenly spaced observations and the power
spectrum of the process must have the form $S(\omega)\propto \omega^{-1}$ to
gain the computational advantage.
This wavelet method has been demonstrated to improve parameter estimation for
transiting exoplanets \citep{Carter:2009}, but these strict requirements make
this method applicable for only a limited set of use cases.

Another GP method that has been used extensively in astronomy is the
hierarchical off-diagonal low rank (HODLR) solver \citep{Ambikasaran:2016}.
This method exploits the fact that many commonly used kernel functions produce
``smooth'' matrices to approximately compute the GP likelihood with the
scaling $\mathcal{O}(N\,\log^2 N)$.
This method has the advantage that, unlike \celerite, it can be used with any
kernel function but, in practice, the cost can still prove to be prohibitively
high for multi-dimensional inputs.
The proportionality constant in the $N\,\log^2N$ scaling of the HODLR method
is a function of the specific kernel and we find~--~using the \project{george}
software package \citep{Foreman-Mackey:2014, Ambikasaran:2016}~--~that this
scales approximately linearly with $J$, but it requires substantial overhead
making it much slower than \celerite\ for all the models we tested.
For large $J \gtrsim 256$ and small $N \lesssim 1000$, we find that
\project{george} can approximately evaluate \celeriteterm\ models with the
same or less computational cost than \celerite, but that \celerite\ is
faster~--~and exact~--~in all other parts of parameter space.

The structured kernel interpolation \citep[SKI/KISS-GP][]{Wilson:2015}
framework is another approximate method that can be used to scale GPs to large
datasets.
This method uses fast interpolation of the kernel into a space where Toeplitz
or Kronecker structure can be used to scale inference and prediction.
The SKI/KISS-GP framework can be applied to scale GP inference with a wide
range of kernels, but the computational cost will depend on the specific
dataset, model, and precision requirements for the approximation.
The SKI method has an impressive $\mathcal{O}(1)$ cost for test-time
predictions and it is interesting to consider how this could be applied to
\celeriteterm\ models.

Many other approximate methods for scaling GP inference exist \citep[see, for
example,][and references therein]{Wilson:2015a} and we make no attempt to make
our discussion exhaustive.
The key takeaway here is that \celerite\ provides an \emph{exact} method for
GP inference for a specific class of one-dimensional kernel functions.
Furthermore, since \celeriteterm\ models can be interpreted as a mixture of
stochastically-driven, damped simple harmonic oscillators, they are a
physically motivated choice of covariance function in many astronomical applications.

\section{Discussion}

Gaussian Process models have been fruitfully applied to many problems in
astronomical data analysis, but the fact that the computational cost generally
scales as the cube of the number of data points has limited their use to
small datasets with $N \lesssim 1000$.
With the linear scaling of \celerite\, we envision that the application of
Gaussian processes will be expanded to the existing and forthcoming large
astronomical time domain surveys such as \kepler\ \citep{Borucki:2010},
\project{K2} \citep{Howell:2014}, \tess\ \citep{Ricker:2014}, \lsst\
\citep{Ivezic:2008}, \project{WFIRST} \citep{Spergel:2015}, and
\project{PLATO} \citep{Rauer:2014}.
Despite the restrictive form of the \celeriteterm\ kernel, with a sufficient
number of components it is flexible enough to describe a wide range of
astrophysical variability.
In fact, the relation of the \celeriteterm\ kernel to the damped,
stochastically-driven harmonic oscillator matches simple models of
astrophysical variability, and makes the parameterization interpretable in
terms of resonant frequency, amplitude, and quality factor.

Our background is in studying transiting exoplanets, a field which has only
recently begun to adopt GP methods for analyzing the noise in stellar light
curves and radial velocity datasets when detecting or characterizing
transiting planets \citep[for example,][]{Carter:2009, Gibson:2012,
Haywood:2014, Barclay:2015, Evans:2015, Rajpaul:2015, Aigrain:2016,
Foreman-Mackey:2016b, Grunblatt:2016, Luger:2016}.
All of these analyses have been limited to small datasets or restrictive
kernel choices, but \celerite\ weakens these requirements by providing a
scalable method for computing the likelihood and a physical motivation for the
choice of kernel.
As higher signal-to-noise observations of transiting exoplanet systems are
obtained, the effects of stellar variability will more dramatically impact the
correct inference of planetary transit parameters.
We predict that \celerite\ will be important for scaling methods of transit
detection \citep{Pope:2016, Foreman-Mackey:2016b}, transit timing
\citep{Agol:2005, Holman:2005}, transit spectroscopy \citep{Brown:2001},
Doppler beaming \citep{Loeb:2003, Zucker:2007}, tidal distortion
\citep{Zucker:2007}, and phase functions \citep{Knutson:2007, Zucker:2007} to
the large astronomical time domain surveys of the future.

\subsection{Other applications and limitations}

Beyond these applications to model stellar variability, the method is
generally applicable to other one-dimensional GP models.
Accreting black holes show time series which may be modeled using a GP
\citep{Kelly:2014}; indeed, this was the motivation for the original technique
developed by Rybicki \& Press \citep{Rybicki:1992, Rybicki:1995}.
This approach may be broadly used for characterizing quasar variability
\citep{MacLeod:2010}, measuring time lags with reverberation mapping
\citep{Zu:2011, Pancoast:2014}, modeling time delays in multiply-imaged
gravitationally-lensed systems \citep{Press:1998}, characterizing
quasi-periodic variability in a high-energy source \citep{McAllister:2016}, or
classification of variable objects \citep{Zinn:2016}.
We expect that there are also applications beyond astronomy.

The \celeriteterm\ formalism can also be used for power spectrum estimation and
quantification of its uncertainties.
A mixture of \celeriteterm\ terms can be used to perform non-parametric
probabilistic inference of the power spectrum despite unevenly-spaced data
with heteroscedastic noise \citep[see][for examples of this
procedure]{Wilson:2013, Kelly:2014}.
This type of analysis will be limited by the quadratic scaling of \celerite\
with the number of terms $J$, but this limits existing methods as well
\citep{Kelly:2014}.

There are many data analysis problems where \celerite\ will not be immediately
applicable.
In particular, the restriction to one-dimensional problems is significant.
There are many examples of multidimensional GP modeling the astrophysics
literature \citep[recent examples from the field of exoplanet characterization
include][]{Haywood:2014, Rajpaul:2015, Aigrain:2016}, where \celerite\ cannot
be used to speed up the analysis.
It is plausible that an extension can be derived to tackle some
multidimensional problems with the right structure~--~simultaneous parallel
time series, for example~--~and we hope to revisit this possibility in future
work.

\newpage
\subsection{Code availability}

Alongside this paper, we have released a well-tested and documented open
source software package that implements the method and all of the examples
discussed in these pages.
This software is available on GitHub
\url{https://github.com/dfm/celerite}\footnote{This version of the paper was
generated with git commit \texttt{\githash} (\gitdate).} and Zenodo
\citep{Foreman-Mackey:2017}, and it is made available under the MIT license.

\acknowledgments
It is a pleasure to thank
Megan Bedell,
Ian Czekala,
Will Farr,
Sam Grunblatt,
David W.\ Hogg,
Dan Huber,
Meredith Rawls,
Dennis Stello,
Jake VanderPlas, and
Andrew Gordon Wilson
for helpful discussions informing the ideas and code presented here.
\response{We would also like to thank the anonymous referee for thorough and
constructive feedback that greatly improved the paper.}

This work was performed in part under contract with the Jet Propulsion
Laboratory (JPL) funded by NASA through the Sagan Fellowship Program executed
by the NASA Exoplanet Science Institute.
EA acknowledges support from NASA grants NNX13AF20G, NNX13A124G, NNX13AF62G,
from National Science Foundation (NSF) grant AST-1615315, and from
NASA Astrobiology Institute's Virtual Planetary Laboratory, supported
by NASA under cooperative agreement NNH05ZDA001C.

This research made use of the NASA \project{Astrophysics Data System} and the
NASA Exoplanet Archive.
The Exoplanet Archive is operated by the California Institute of Technology,
under contract with NASA under the Exoplanet Exploration Program.

This paper includes data collected by the \kepler\ Mission. Funding for the
\kepler\ Mission is provided by the NASA Science Mission directorate.
We are grateful to the entire \kepler\ team, past and present.
These data were obtained from the Mikulski Archive for Space Telescopes
(MAST).
STScI is operated by the Association of Universities for Research in
Astronomy, Inc., under NASA contract NAS5-26555.
Support for MAST is provided by the NASA Office of Space Science via grant
NNX13AC07G and by other grants and contracts.

This research made use of Astropy, a community-developed core Python package
for Astronomy \citep{Astropy-Collaboration:2013}.

\facility{Kepler}
\software{%
     \project{AstroPy} \citep{Astropy-Collaboration:2013},
     \project{corner.py} \citep{Foreman-Mackey:2016},
     \project{Eigen} \citep{Guennebaud:2010},
     \project{emcee} \citep{Foreman-Mackey:2013},
     \project{george} \citep{Ambikasaran:2016},
     \project{Julia} \citep{Bezanzon:2012},
     \project{LAPACK} \citep{Anderson:1999},
     \project{matplotlib} \citep{Hunter:2007},
     \project{numpy} \citep{Van-Der-Walt:2011},
     \project{transit} \citep{Foreman-Mackey:2016a},
     \project{scipy} \citep{Jones:2001}.
}

\vspace{5ex}
\appendix

\section{Ensuring positive definiteness for celerite models}\sectlabel{psd}

For a GP kernel to be valid, it must produce a positive definite covariance
matrix for all input coordinates.
For stationary kernels, this is equivalent~--~by Bochner's theorem \citep[see
Section 4.2.1 in][]{Rasmussen:2006}~--~to requiring that the kernel be the
Fourier transform of a positive finite measure.
This means that the power spectrum of a positive definite kernel must be
positive for all frequencies.
This result is intuitive because, since the power spectrum of a process is
defined as the expected squared amplitude of the Fourier transform of the time
series, it must be non-negative.

Using \eq{celerite-psd}, we find that for a single \celeriteterm\ term, this
requirement is met when
\begin{eqnarray}\eqlabel{psd-req}
\frac{(a_j\,c_j+b_j\,d_j)\,({c_j}^2+{d_j}^2)+(a_j\,c_j-b_j\,d_j)\,\omega^2}
{\omega^4+2\,({c_j}^2-{d_j}^2)\,\omega^2+{({c_j}^2+{d_j}^2)}^2} > 0 \quad.
\end{eqnarray}
The denominator is positive for all $c_j \ne 0$ and it can be shown that, when
$c_j=0$, \eq{psd-req} is satisfied for all $\omega \ne d_j$ where the power is
identically zero.
Therefore, when $c_j \ne 0$, we require that the numerator is positive for all
$\omega$.
This requirement can also be written as
\begin{eqnarray}
a_j\,c_j > -b_j\,d_j \\
a_j\,c_j > b_j\,d_j \quad.
\end{eqnarray}
Furthermore, we can see that $a_j$ must be positive since $k(0) = a_j$ should
should be positive and, similarly, by requiring the covariance to be finite at
infinite lag, we obtain the constraint $c_j \ge 0$.
Combining these results, we find the constraint
\begin{eqnarray}
    |b_j\,d_j| < a_j\,c_j \quad.
\end{eqnarray}

\response{The constraint for $J > 1$ is more complicated so, in most cases, we
require that each term is positive definite using this relationship and use
the fact that a product of sum of positive definite kernels will also be
positive definite \citep{Rasmussen:2006} to show that the full model is
positive.}
However, in the case of $J$ general \celeriteterm\ terms, we can check for
negative values of the PSD by solving for the roots of the power spectrum;
if there are any real, positive roots, then the power-spectrum goes negative
(or zero), and thus does not represent a valid kernel.
We rewrite the power spectrum, \eqalt{celerite-psd}), abbreviating with
$z = \omega^2$:
\begin{equation}
S(\omega)=  \sum_{j=1}^J \frac{q_j\,z + r_j}{z^2+s_j\,z + t_j} = 0
\end{equation}
where
\begin{eqnarray}
q_j &=& a_j\,c_j-b_j\,d_j\\
r_j &=& (d_j^2+c_j^2)(b_j\,d_j+a_j\,c_j)\\
s_j &=& 2(c_j^2-d_j^2)\\
t_j &=& (c_j^2+d_j^2)^2.
\end{eqnarray}
The denominators of each term are positive, so we can multiply through by
$\prod_j \left(z^2+s_jz + t_j\right)$ to find
\begin{equation}
Q_0(z) = \sum_{j=1}^J (q_j z + r_j)\prod_{k \ne j}\left(z^2+s_kz +
    t_k\right) = 0\quad,
\end{equation}
which is a polynomial with order $2\,(J-1)+1$.
With $J=2$, this yields a cubic equation whose roots can be obtained exactly.

For arbitrary $J$, a procedure based upon Sturm's theorem \citep{Dorrie:1965}
allows one to determine whether there are any real roots within the range
$(0,\,\infty]$.
We first construct $Q_0(z)$ and its derivative $Q_1(z) = {Q_0}^\prime(z)$,
and then loop from $k=2$ to $k=2\,(J-1)+1$, computing
\begin{eqnarray}
Q_k(z) = -{\rm rem}(Q_{k-2},\,Q_{k-1})
\end{eqnarray}
where the function ${\rm rem}(p,\,q)$ is the remainder polynomial after
dividing $p(z)$ by $q(z)$.

We evaluate the coefficients of each of the polynomials in the series by
evaluating $f_0 = \{Q_0(0),\,\ldots,\,Q_{2\,(J-1)+1}(0)\}$ to give us the signs
of these polynomials evaluated at $z=0$.
Likewise, we evaluate the coefficients of the largest order term in each
polynomial that gives the sign of the polynomial as $z \rightarrow \infty$,
$f_\infty$.
With the sequence of coefficients $f_0$ and $f_\infty$, we then determine how
many times the sign changes in each of these, where $\sigma(0)$ is the number
of sign changes at $z=0$, and $\sigma(\infty)$ is the number of sign changes
at $z \rightarrow \infty$.
The total number of real roots in the range
$(0,\,\infty]$ is given by $N_{+}=\sigma(0)-\sigma(\infty)$.

We have checked that this procedure works for a wide range of parameters, and
we find that it robustly matches the number of positive real roots which we
evaluated numerically.
The advantage of this procedure is that it does not require computing the
roots, but only carrying out algebraic manipulation of polynomials to
determine the number of positive real roots.
If a non-zero real root is found, the likelihood may be set to zero.

\response{
\section{Semiseparable matrix operations} \sectlabel{operations}

In this appendix, we summarize some algorithms for manipulating semiseparable
matrices and discuss how these can be used with GP models and \celerite.

\subsection{Multiplication}

The product of a rank-$R$ semiseparable matrix with general vectors and
matrices can be computed in $\mathcal{O}(N\,R)$ operations
\citep{Vandebril:2007}.
In the context of \celeriteterm\ models, na\"ive application of these methods
will result in numerical overflow and underflow issues, much like we found
with the Cholesky factorization.
To avoid these errors, we derive the following numerically stable algorithm
for computing the dot product of a \celeriteterm\ covariance matrix
\begin{eqnarray}
\bvec{y} = K\,\bvec{z}
\end{eqnarray}
using the semiseparable representation for $K$ defined in \eqalt{reparam}:
\begin{eqnarray}
f^+_{n,j} &=& \phi_{n+1,j}\,\left[f^+_{n+1,j} +
    \tilde{U}_{n,j}\,z_{n+1}\right] \\
f^-_{n,j} &=& \phi_{n,j}\,\left[f^-_{n-1,j} +
    \tilde{V}_{n-1,j}\,z_{n-1}\right] \\
y_n &=& A_{n,n}\,z_n + \sum_{j=1}^{2\,J} \left[ \tilde{V}_{n,j} f^+_{n,j} +
    \tilde U_{n-1,j} f^-_{n,j}\right] \quad,
\end{eqnarray}
where $f^+_{N,j} = 0$ and $f^-_{1,j} = 0$.
This algorithm requires two sweeps~--~$n=2,\,\ldots,\,N$ to compute $f^-$ and
$n=N-1,\,\ldots,\,1$ to compute $f^+$~--~with a scaling of $\mathcal{O}(N\,J)$.

\subsection{Sampling data from a celerite process}

To sample a dataset $\bvec{y}$ from a GP model for fixed parameters
$\bvec{\theta}$ and $\bvec{\alpha}$, we must compute
\begin{equation} \eqlabel{sim}
\bvec{y} = \bvec{\mu}_\bvec{\theta}(\bvec{t})+
    K_\bvec{\alpha}^{1/2}\,\bvec{q}
\end{equation}
where $\bvec{\mu}_\bvec{\theta}(\bvec{t})$ is the mean function evaluated at
the input coordinates $\bvec{t}$, $\bvec{q}$ is a vector of draws from the
unit normal $q_i \sim \mathcal{N}(0,\,1)$, and
\begin{eqnarray}
K_\bvec{\alpha} = K_\bvec{\alpha}^{1/2} \,
    {\left(K_\bvec{\alpha}^{1/2}\right)}^T\quad.
\end{eqnarray}
For a general matrix $K_\bvec{\alpha}$, the cost of this operation scales as
$\mathcal{O}(N^3)$ to compute $K_\bvec{\alpha}^{1/2}$ and $\mathcal{O}(N^2)$
to compute the dot product.
For \celeriteterm\ models, we use the semiseparable representation of the
Cholesky factor that we derived in \sect{cholesky} to compute the dot product
\eq{sim} in $\mathcal{O}(N\,J)$.
Using the notation from \sect{scalable}, the algorithm to compute this dot
product is
\begin{eqnarray}
f_{n,j} &=& \phi_{n,j}\,\left[f_{n-1,j} + \tilde{W}_{n-1,j}\,
    \sqrt{D_{n-1,n-1}}\,q_{n-1}\right]\\
y_n &=& \sqrt{D_{n,n}}\,q_n + \sum_{j=1}^{2\,J} \tilde{U}_{n,j}\,f_{n,j}
\end{eqnarray}
where $f_{1,j} = 0$ for all $j$.
This scalable algorithm is useful for generating large simulated
datasets for experiments with \celerite\ and for performing posterior
predictive checks.

\subsection{Interpolation \& extrapolation}

The predictive distribution of a GP at a vector of $M$ input coordinates
\begin{eqnarray}
\bvec{y}^* = \left(\begin{array}{ccccc}
    t_1^*\quad && \cdots\quad && t_M^*
\end{array}\right)^\T
\end{eqnarray}
conditioned on a dataset $(\bvec{t},\,\bvec{y})$ and GP
parameters $(\bvec{\theta},\,\bvec{\alpha})$, is a normal $\bvec{y}^* \sim
\mathcal{N}(\bvec{\mu}^*,\,K^*)$ with mean
\begin{eqnarray}\eqlabel{pred-mean}
\bvec{\mu}^* &=& \bvec{\mu}_\bvec{\theta}(\bvec{t}^*) +
    K(\bvec{t}^*,\,\bvec{t})\,{K(\bvec{t},\,\bvec{t})}^{-1}\,\left[\bvec{y} -
    \bvec{\mu}_\bvec{\theta}(\bvec{t})\right]
\end{eqnarray}
and covariance
\begin{eqnarray}
K^* &=& K(\bvec{t}^*,\,\bvec{t}^*) -
    K(\bvec{t}^*,\,\bvec{t})\,{K(\bvec{t},\,\bvec{t})}^{-1}\,
    K(\bvec{t},\,\bvec{t}^*)
\end{eqnarray}
where $K(\bvec{v},\,\bvec{w})$ is the covariance matrix computed between
\bvec{v} and \bvec{w} \citep{Rasmussen:2006}.

Na\"ively, the computational cost of computing \eq{pred-mean} for a
\celeriteterm\ model scales as $\mathcal{O}(N\,M)$ if we reuse the Cholesky
factor derived in \sect{cholesky}.
It is, however, possible to improve this scaling to $\mathcal{O}(n\,N +
m\,M)$ where $n$ and $m$ are integer constants.
To derive this, we expand \eq{pred-mean} using \eq{celerite-kernel}
\begin{eqnarray}
\mu^*_m &=& K(t^*_m,\bvec{t}) \bvec{z} \\
    &=& \sum_{n=1}^N \sum_{j=1}^J e^{-c_j \vert t^*_m-t_n\vert}\,\left[
        a_j\,\cos{(d_j\,\vert t^*_m-t_n\vert)} +
        b_j\,\sin{(d_j\,\vert t^*_m-t_n\vert)}
    \right] \,z_{n} \quad. \nonumber
\end{eqnarray}
where
\begin{eqnarray}
\bvec{z} = {K(\bvec{t},\,\bvec{t})}^{-1}\,\left[\bvec{y} -
    \bvec{\mu}_\bvec{\theta}(\bvec{t})\right] \quad.
\end{eqnarray}
Dividing the sum over $n$ into $\{t_1,...,t_{n_0}\} < t^*_m$ and
$\{t_{n0+1},...,t_N\} \ge t^*_m$, gives
\begin{eqnarray}
\mu^*_m &=& \sum_{j=1}^J\sum_{n=1}^{n_0}
    e^{-c_j\,(t^*_m-t_n)}\,\left[a_j\,\cos{(d_j\,(t^*_m-t_n))}
        +b_j\,\sin{(d_j\,(t^*_m-t_n))}\right]\,z_n \\
&+& \sum_{j=1}^J\sum_{n=n_0+1}^N  e^{-c_j\,(t_n-t^*_m)}\,\left[
    a_j\,\cos{(d_j\,(t_n-t^*_m))}+b_j\,\sin{(d_j\,(t_n-t^*_m))}\right]\,z_n
\quad. \nonumber
\end{eqnarray}
We compute this using two passes, a forward pass $n_0 = 1,\,\ldots,\,N$ and
a backward pass $n_0 = N,\,\ldots,\,1$.
Defining
\begin{eqnarray}
Q^-_{n,k} &=& \left[Q^-_{n-1,k}+z_n\,\tilde V_{n,k}\right]\,e^{-c_{k//2}\,(t_{n+1}-t_{n})},\\
X^-_{m,n,k} &=& e^{-c_{k//2}\,(t_m^*-t_{n+1})}\,\tilde U_{m,k}^*,\\
Q^+_{n,k} &=& \left[Q^+_{n+1,k}+z_n\,\tilde U_{n,k}\right]\,e^{-c_{k//2}(t_{n}-t_{n-1})},\\
X^+_{m,n,k} &=& e^{-c_{k//2}\,(t_{n-1}-t_m^*)}\,\tilde V_{m,k}^* \quad,
\end{eqnarray}
where $t_0=t_1$, $t_{N+1}=t_N$, $Q^-_{0,k} = 0$ and $Q^+_{N+1,k}=0$ for $k=1$
to $2\,J$, and in $X^\pm_{m,n,k}$ the expressions
for $\tilde U_{m,i}^*$ and $\tilde V_{m,i}^*$ are evaluated at $t_m^*$.
For each value of $m$, $Q^\pm$ are recursively updated over $n$ until $n_0$ is
reached, at which point the predicted mean can be computed from
\begin{equation}
\mu^*_m = \sum_{k=1}^{2\,J}
    \left[Q^-_{n_0,k}X^-_{m,n_0,k}+Q^+_{n_0+1,k}X^+_{m,n_0+1,k}\right] \quad.
\end{equation}
}

\bibliography{celerite}

\end{document}